\documentclass[twocolumn,showpacs,preprintnumbers,amsmath,amssymb,amsfonts,floatfix]{revtex4}
\usepackage{booktabs}
\usepackage{upgreek}
\usepackage{CJK}
\usepackage{lipsum,graphicx}

\usepackage{csquotes} 
\usepackage{caption}
\usepackage{lipsum,graphicx}
\usepackage{multirow}
\usepackage{amsfonts}
\usepackage{amsmath}
\usepackage{amssymb}

\usepackage{float}
\usepackage{graphicx}
\usepackage{dcolumn}
\usepackage{bm}
\usepackage[section] {placeins}
\usepackage{lipsum}
\usepackage{color}
\usepackage{amsmath}
\usepackage{lipsum}

\usepackage{titlesec}
\titlespacing*{\section}{0pt}{1.1\baselineskip}{\baselineskip}

\usepackage{pdfpages}
\usepackage{natbib}
\begin{document}

\newcommand {\nc} {\newcommand}
\nc {\IR} [1] {\textcolor{red}{#1}} 
\nc {\IB} [1] {\textcolor{blue}{#1}}
\begin{CJK}{UTF8}{gbsn}
\title{Non--local interactions in the $(d,p)$ surrogate method for $(n,\gamma)$ reactions}
\author{
Weichuan Li (李纬川)$^{1,2}$, Gregory Potel$^{1}$ and Filomena Nunes$^{1,3}$ \\
$^{1}$National Superconducting Cyclotron Laboratory,
Michigan State Univeristy,
East Lansing, Michigan 48824, USA\\
$^{2}$Department of Physics and Astronomy,
Rutgers Univeristy,
New Brunswick, New Jersey 08901, USA\\
$^{3}$Department of Physics and Astronomy,
Michigan State Univeristy,
East Lansing, Michigan 48824, USA}
\date{\today}

\begin{abstract}

	\begin{description}
		\item[Background] Single-neutron transfer reactions populating states in the continuum are interesting both for structure and astrophysics. In their description often global optical potentials are used for the nucleon-target interactions, and these interactions are typically local. In our work, we study the effects of nonlocality  in $(d,p)$ reactions populating continuum states. This work is similar to that of ~\cite{Titus_prc2014} but now for transfer to the continuum.
		\item[Purpose]A theory for computing cross sections for inclusive
		processes $A (d,p) X$ was explored in~\cite{Potel_PRC2015}. Therein, local optical potentials were used to describe the nucleon-target effective interaction. The goal of the present work is to extend the theory developed in~\cite{Potel_PRC2015} to investigate the effects of including nonlocality in the effective interaction on the relevant reaction observables. 
		\item[Method] We implement the R--matrix method to solve the non--local equations both for the nucleon wavefunctions and the propagator. We then apply the method to systematically study the inclusive process of $(d,p)$ on $^{16}$O, $^{40}$Ca, $^{48}$Ca and $^{208}$Pb at 10, 20 and 50 MeV. We compare the results obtained when non--local interactions are used with those obtained when local equivalent interactions are included.
		\item[Results] We find that nonlocality affects different pieces of the model in complex ways. The competition between the reduction of the propagator and the neutron wavefunction in the region of interest and the increase of the magnitude of the interaction produces varying effects on the cross section. Depending on the beam energy and the target, the non-elastic breakup can either increase or decrease. Effects on the heavier targets can be as large as $85$\%.
		\item[Conclusions] While the non-elastic transfer cross section for each final spin state can change considerably, the main prediction of the model ~\cite{Potel_PRC2015}, namely the shape of the spin distributions, remains largely unaltered by nonlocality.
	\end{description}
	
\end{abstract}

\maketitle

\section{Motivation}
Neutron capture reactions are an essential piece of the puzzle in understanding the formation of nuclei  heavier than $^{56}$Fe ~\cite{Spyrou_JPG2017}. One of the major efforts in our community is understanding the $r$-process~\cite{Mumpower_PPNP2016} which requires neutron capture rates for short-lived isotopes. Neutron capture reactions are also relevant to new reactor designs and for stockpile stewardship~\cite{Cizewski_NIM2007,Cizewski_EPJ2017}. For many of these applications, the isotopes are sufficiently far from stability that it is impossible to measure $(n,\gamma)$ directly and indirect methods need to be used. 

This work is focused on the use of $(d,p)$ reactions as a surrogate for $(n,\gamma)$. 
Many of the isotopes in the $r$-process path are unstable but not close to the neutron drip line. For these isotopes, resonant capture is the dominant capture process. In resonant neutron capture, the neutron populates compound excited states in the continuum, and then decays through $\gamma$-ray or particle emission. Any indirect method for $(n,\gamma)$ assumes that it populates the same compound states. In the $(d,p)$ surrogate method, it is assumed that the deuteron will transfer its neutron onto the target, forming the same compound nucleus that is formed in the direct $(n,\gamma)$, so that, by measuring the $\gamma$-decay following the $(d,p)$ reaction, one can constrain the neutron capture of interest. However, one does need to rely on reaction theory for a reliable extraction~\cite{Carlson_2015,Jin_PRC2015,Jin_PRC2015_2,Potel_PRC2015,Potel_EPJA2017}.

There are some known complications to the extraction. The model described in~\cite{Potel_PRC2015} relies on a 2-step process, first the deuteron breaks up and then the neutron gets captured (absorbed) by the target.  But, in this model, there is some probability that the neutron will not be absorbed, leading to elastic breakup (EB). The EB cross section needs to be subtracted from the total $(d,p)$ cross section. The remaining component, usually called the non-elastic breakup (NEB), connects with the compound nucleus formation in $(n,\gamma)$ reactions. One of the most important inputs that theory provides in the analysis of the $(d,p)$ surrogate data are the spin distributions, essential for correcting the  weights  of the various final spins to enable a reliable extraction of the $(n,\gamma)$ cross section.   It turns out that the spin distribution obtained through $(d,p)$ is not the same as that obtained in $(n,\gamma)$. An example of a successful application of the method can be found in~\cite{Potel_PRC2018}.

Up to now, the models of $(d,p)$ surrogate reactions~\cite{Carlson_2015,Jin_PRC2015,Jin_PRC2015_2,Potel_PRC2015,Potel_EPJA2017} have used local nucleon-target optical potentials. However, 
it is well known that the nucleon-nucleus effective interactions are non--local due to the channel couplings and antisymmetrization of many-body wavefunctions (e.g.~\cite{Canton_PRL2005,Fraser_EPJ2008,Rawitscher_PRC1994,Feshbach_AP1958,Feshbach_AP1962}). In the last decade, there has been a considerable amount of effort to develop microscopic optical potentials which are intrinsically non--local. An example is the dispersive optical model (DOM) theory which was first introduced by Mahaux and Sartor~\cite{Mahaux1991} and has been implemented and applied by the St. Louis group~\cite{Mahzoon_PRL2014}. Another approach is to obtain the optical potential using ab initio methods. An example of a recent effort using Coupled Cluster theory is discussed in~\cite{Rotureau_PRC2017}. Although the work in~\cite{Rotureau_PRC2017} demonstrates that ab initio optical potentials are still not of practical use, it is important to incorporate, in the implementation of the reaction theory for surrogates, the capability of dealing with non--local interactions.  Furthermore, it is crucial to study the effects of nonlocality in the predictions for the surrogate process to determine whether future analyses need to take this aspect into account.

A study of the effects of nonlocality on reaction observables has a long history, including the well known works of Austern~\cite{Austern_PRB1965,Austern_book} and Fiedeldey~\cite{Fiedeldey_NP1966} (codes like TWOFNR~\cite{lgarashi_code} and DWUCK~\cite{DWUCK_1969} contain such non--local corrections) . This topic has been recently revisited~\cite{Titus_prc2014,Titus_PRC2016}.
In~\cite{Titus_prc2014} nonlocality was only included in the nucleon channels, both in calculating the neutron bound-state wavefunction and the proton distorted wave.
Nonlocality had opposing effects on these two wavefunctions. The largest effect was the reduction of the neutron wavefunction in the interior which, due to normalization, produced an increase of the peripheral part. This resulted in an overall increase of the transfer cross section when non--local interactions were included.
The method was then generalized to include nonlocality in the deuteron channel~\cite{Titus_PRC2016}. Whether using the distorted wave approximation (DWBA) or the adiabatic wave approximation (ADWA),  nonlocality was found to have a considerable effect on transfer angular distributions, mostly affecting the magnitude but sometimes also the shape. Given these results, one might expect similar effects in $(d,p)$ reactions populating continuum states. Here we will use the non--local optical potential derived by Perey and Buck~\cite{Perey_NP1962} to make it easier to compare with previous work~\cite{Titus_prc2014,Titus_PRC2016}. However, it is understood that the {\it actual} non--local optical potential from nature is not simply of Gaussian form and should still contain some energy dependence~\cite{lovell_prc2017}. The energy dependence is usually interpreted as a consequence of the coupling with collective states, and it tends to give rise to a longer--range non--locality.

The work performed in~\cite{Titus_prc2014,Titus_PRC2016} made use of an iterative method for solving the non--local Sch\"odinger equation~\cite{Titus_CPC2016}. However this method is not very efficient and this became a serious issue  when considering transfer to the continuum which involves the computation of a whole range of neutron scattering final states. For this reason, and in order to complete this project, we  implemented a modified R--matrix method following the work by Descouvemont and Baye ~\cite{Descouvemont_RPP2010}. Transforming the finite differences problem into a linear algebra problem provided three orders of magnitude improvement in computing time for the calculations involving non--local interactions ~\cite{bonitati}. 

Following the systematic study of~\cite{Titus_prc2014,Titus_PRC2016}, we study the non-elastic breakup $A (d,p) X$ on $^{16}$O, $^{40}$Ca, $^{48}$Ca and $^{208}$Pb, at various beam energies (10, 20 and 50 MeV). The results obtained using non--local optical potentials are compared with those obtained using local equivalent potentials (LEP). We look at the direct NEB term (Udagawa term~\cite{Udagawa_PRL1980}) as well as the non--orthogonality term (Hussein and McVoy term~\cite{Hussein_NPA1985}). We also consider the spin distributions in detail.

In section II, we provide a brief theoretical background, including a summary of the formalism of~\cite{Potel_PRC2015} and the details of the R--matrix method as we implemented it. Next, in Section III we provide the numerical details associated with our calculations. This is followed by the results in Section IV. The conclusions are drawn in Section VI.

\section{Theoretical background}
\subsection{Inclusive $(d,p)$ reactions}
\begin{figure}[t!]
	
	\includegraphics[width=.45\textwidth,height=7.8cm]{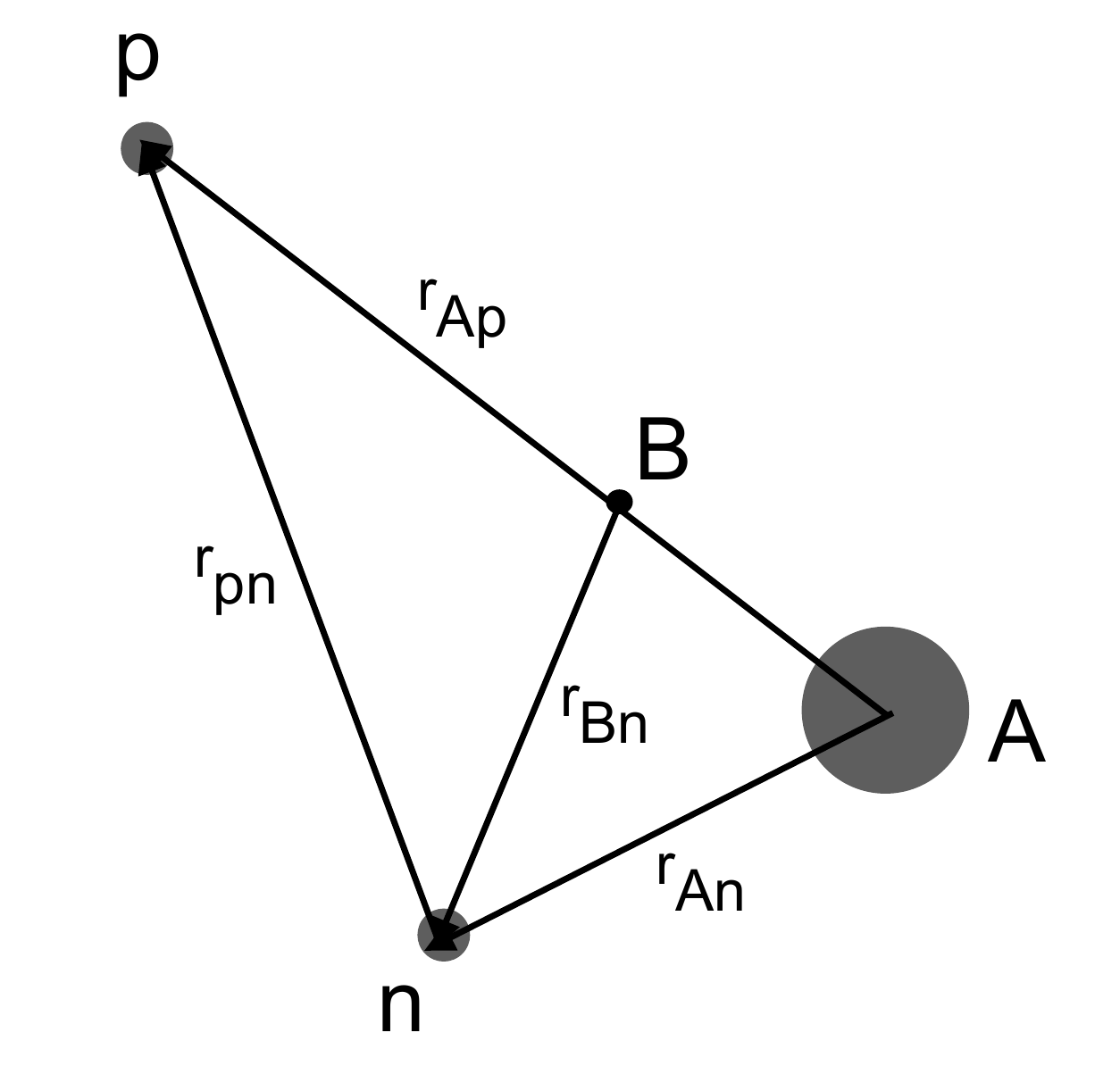}
	\caption{Schematic representation of the coordinates used in this work.} 
	\label{fig:coordinate} 
\end{figure}

In this work we describe the NEB channels populated in $A (d,p) X$  processes within a two--step mechanism. The first step consists in the breaking up of the deuteron  due to the action of the neutron--target potential, while, in the second step, the released neutron interacts with the target.  We adopt the system of coordinates shown in Fig.~\ref{fig:coordinate} where $B$ represents a state of the $A+n$ system. The neutron--target interaction is described by the single--particle Green's function  $G^{\text{opt}}_{B}$, and results in the population of the elastic breakup and NEB final channels (the superscript $^{opt}$ is meant to indicate that the single--particle operator $G^{\text{opt}}_B$ has been obtained by \textit{optical reduction} of the many--body propagator $G_B$, see~\cite{Potel_PRC2015}). For low neutron--target relative energies ($\lesssim 1$ MeV),  the compound nucleus formation dominates the NEB processes, and $(d,p)$ reactions can be used as surrogate for ($n,\gamma$) reactions. 

For negative neutron energies, and in the limit of the imaginary part of the neutron--target interaction going to zero, the NEB contribution can be shown to reproduce the standard DWBA with a neutron wavefunction generated by the real part of this interaction~\cite{Potel_PRC2015}. By focusing our attention on the NEB component of the $A (d,p) X$ reaction, we can then consider this work as a natural extension of~\cite{Titus_prc2014} to positive energies, as well as describing the channel of interest for the $(d,p\gamma)$ surrogate method.  

The nonelastic-breakup differential cross section can be expressed as~\cite{Potel_PRC2015}
 \begin{equation}
 \begin{split}
 \left.\frac{d^{2} \sigma}{d E_{p} d \Omega_{p}}  \right]^{\text{NEB}} & = \left.\frac{d^{2} \sigma}{d E_{p} d \Omega_{p}}  \right]^{\text{UT}} +\left.\frac{d^{2} \sigma}{d E_{p} d \Omega_{p}}  \right]^{\text{HM}}  \\& +\frac{2\pi}{\hbar v_{d}} \rho_{p}(E_{p}) \left(\right.2 \text{Re} \left \langle \Phi_{n}^{\text{HM}} |W_{An}| \Phi_{n}  \rangle \right). 
 \end{split}
 \label{equ:cross_section}
 \end{equation}
 In order to understand the various terms in Eq.~(\ref{equ:cross_section}) it is necessary to introduce some important ingredients of the theory. These include the proton level density:
   \begin{equation}
 \rho_{p}(E_{p})=\frac{m_{p}k_{p}}{8\pi^{3}\hbar^{2}}
 \end{equation}  
 where $E_{p}$ describes the kinetic energy of proton,
 and $W_{An}$ is the imaginary part  of the neutron--target potential, which  accounts for all reaction channels.  The first term in Eq.~(\ref{equ:cross_section}) describes the direct (Udagawa--Tamura, UT~\cite{Udagawa_PRL1980}) process, and can be written as 
 	\begin{equation}
 	\begin{split}
 	\left.\frac{d^{2} \sigma}{d E_{p} d \Omega_{p}}  \right]^{\text{UT}} & = \frac{2\pi}{\hbar v_{d}} \rho_{p}(E_{p}) \left( \text{Im}\left \langle \Phi_{n} |W_{An}| \Phi_{n} \right \rangle \right),
 	\end{split}
 	\label{equ:cross_section_UT}
 	\end{equation} 
 	while the second term corresponds to the  nonorthogonality (Hussein--McVoy, HM) contribution ~\cite{Hussein_NPA1985},
 
 	\begin{equation}
 	\begin{split}
 	\left.\frac{d^{2} \sigma}{d E_{p} d \Omega_{p}}  \right]^{\text{HM}} & = \frac{2\pi}{\hbar v_{d}} \rho_{p}(E_{p}) \left( \left \langle \Phi_{n}^{\text{HM}} |W_{An}| \Phi_{n}^{\text{HM}} \right \rangle \right).
 	\end{split}
 	\label{equ:cross_section_HM}
 	\end{equation}
 The third term corresponds to the interference between the UT and HM contributions, testifying to fact that both terms contribute coherently to the total NEB cross section. 

 An important element in the theory is the neutron wavefunction, after breakup, in the field of the target:
  \begin{equation}
  \Phi_{n}=G_{B}^{\text{opt}} S.
  \label{equ:neutronwave}
  \end{equation}
This $\Phi_n$ is the result of the application of the single--particle Green's function (the propagator) to the source term
  \begin{equation}
  S=\left(\chi_{p} |\left(U_{Ap}-U_{Ad}+U_{An}\right)| \chi_{d} \phi_{d}
  \right \rangle. 
  \label{equ:source}
  \end{equation}
 This source term accounts for the direct breakup of the deuteron, the first step in our two--step description. The deuteron distorted wave $\chi_d$ is generated by the local optical potential $U_{Ad}$, while  the proton  scattering wavefunction $\chi_{p}$ is a positive--energy solution of a Schr\"odinger equation with the non--local potential $U_{Ap}$ (the round bracket on the left indicates integration over the proton coordinates only). Note that the neutron--target interaction $U_{An}$ is also, in general, non--local. 

The non--orthogonality (Hussein--McVoy, HM) function is defined as
\begin{equation}
\Phi_{n}^{\text{HM}}=\left ( \chi_{p}|\phi_{d} \chi_{d}\right \rangle,
\end{equation}
and accounts for the fact that, in prior form, there is an additional correction to the Udagawa-Tamura term ~\cite{Udagawa_PRL1980}.

In order to derive the Green's function $G_{B}^{\text{opt}}$ and the scattering wavefunction $\chi_{p}$  with non--local potentials, we have implemented dedicated numerical methods. We will briefly address them in the following sections. An extended description of this method will appear elsewhere~\cite{method-paper}.

\subsection{R--matrix method}
The proton scattering wavefunction $\chi_p$ is obtained by solving the non--local Schr\"odinger equation 
\begin{equation}
(T_{l_{p}}-E)\chi^{l_{p}}_{p}(r)=\int_{0}^{\infty}\chi^{l_{p}}_{p}(r')U_{Ap}(r,r')r'^2dr',
\label{equ:schrodinger_nonlocal}
\end{equation}
where
\begin{equation}
T_{l_{p}}=\frac{-\hbar^{2}}{2\mu} \left[\frac{d^{2}}{dr^{2}}-\frac{l_{p}(l_{p}+1)}{r^{2}}\right],
\end{equation}
and $\chi^{l_{p}}_{p}(r)$ is the $l_{p}$--component of the standard partial wave decomposition of $\chi_p(\mathbf r)$. 
This equation has been solved with an iterative method by Titus et al.~\cite{Titus_CPC2016}. Here, we propose  a faster and more robust method based on an R--matrix expansion~\cite{Descouvemont_RPP2010}.

The R--matrix method was first introduced by Wigner with the aim of describing resonances~\cite{Wigner_PR1946}, and it has been developed and applied widely in atomic and nuclear physics~\cite{Lane_RMP1958,Baye_JPB1998,Light_JCP1976}. The basic idea of the R--matrix method is to divide the configuration space into an internal ($r<a$)  and an external $(r>a)$ region by defining a suitable radius $a$. This limiting radius $a$ should be large enough so that  the nuclear part of the interaction $U_{Ap}$ is negligible for $r>a$. In practice, we check that the results are independent of the choice of $a$.  The boundary conditions can be conveniently enforced by defining the Bloch operator
\begin{equation}
\mathcal{L}=\frac{\hbar^{2}}{2\mu} \delta(r-a)\frac{d}{dr},
\end{equation}
and solving the associated integro--differential equation
\begin{multline}
(T_{l_{p}}+\mathcal{L}-E)\chi^{int,l_{p}}_{p}(r)+\int_{0}^{a}U_{Ap}(r,r')\chi^{int,l_{p}}_p(r')r'^2dr'\\=\mathcal{L}\, \chi_{p}^{ext,l_{p}}(r).
\label{equ:bloch}
\end{multline}
The external $(r>a)$ wavefunction $\chi_{p}^{ext,l_{p}}(r)$ is  a  Coulomb wave:
\begin{equation}
\chi_{p}^{ext,l_p}(r)\underset{r\rightarrow\infty}{\rightarrow} i/2\left[ H^-_{l_p} (\eta,kr)+ \exp (2i\delta_{l_p})H^+_{l_p}(\eta,kr) \right],
\label{equ:ext1}
\end{equation}
where $\delta_{l_p}$ is the phase shift for partial wave $l_p$, $\eta$ the Sommerfeld parameter, and $k$ is the proton wave number.
The internal ($r<a$) wave function is expanded in the Lagrange basis:
\begin{equation}\label{Lexpansion}
\chi^{int,l_{p}}(r)=\sum_{j=1}^{N}c_{j} \varphi_{j}(r).
\end{equation}	
The Lagrange functions $\varphi_{i}(r)$ are defined in the interval $(0,a)$ as
\begin{equation}
\varphi_{i}(r)=(-)^{N+i}\frac{r}{ax_{i}} \sqrt{ax_{i}(1-x_{i})}\frac{P_{N}(2r/a-1)}{r-ax_{i}},
\label{equ:lagrange}
\end{equation}
where $P_{N}$(x) is a Legendre polynomial of degree $N$, and the points $x_{i}$ verify
\begin{equation}
P_{N}( 2x_{i} -1 ) = 0 .
\end{equation}
 The factor $r/ax_{i}$ enforces the regular behavior of  the basis functions at the origin. The next step is to compute the Hamiltonian matrix in the Lagrange basis,
\begin{equation}
C_{ij}(E)= \langle \varphi_{i}| T_{l_{p}} +\mathcal{L} + U_{Ap}-E| \varphi_{j}\rangle.
\end{equation}
In the Lagrange basis, the local and non--local matrix elements present in the equation above are particularly easy to compute, without the need to perform any numerical integration (see \cite{Descouvemont_RPP2010}).
The integro--differential equation (\ref{equ:bloch}) is thus recast into a linear algebra problem that can be solved for the $c_i$ coefficients of the expansion (\ref{Lexpansion}), 
\begin{equation}
\sum_{ij=1}^{N}C_{ij}(E) c_{j}= \sum_{i=1}^{N}\frac{\hbar^{2}}{2\mu } \chi_{p}^{ext,l_{p}}(a)\varphi_{i}(a).
\end{equation} 
It is to be noted that this method can also be used to compute bound states, in which case the Coulomb wave functions enforcing the boundary conditions need be replaced by Whittaker functions.

\subsection{Computing the Green's function}
The target--neutron Green's function G$^{\text{opt}}_{B}$ can be written as
\begin{equation}
G_{l} (r_{Bn},r_{Bn}')=\frac{\chi_{l}(k_{n},r_{<Bn})g_{l}(k_{n},r_{>Bn})}{\mathcal W(r_{Bn}')r_{Bn}r_{Bn}'},
\label{equ:green1}
\end{equation}
where $r_{<Bn}$ stands for $\min\{r_{Bn},r_{Bn}'\}$ and $r_{>Bn}$ stands for $\max\{r_{Bn},r_{Bn}'\}$. Note that the Wronskian  $\mathcal W=\chi'_l\,g_{l}-\chi_l\,g_{l}'$ is $r_{Bn}$--dependent when the potential is non--local. The function $\chi_{l}(k_{n},r_{<Bn})$ $(g_{l}(k_{n},r_{>Bn}))$ is the regular (irregular) solution of the non--local Schr\"odinger equation: 
\begin{equation}
( T_{l} -U_{An}(r_{Bn},r'_{Bn})+E)\left\{\chi_{l}(k_{n},r_{<Bn}),g_{l}(k_{n},r_{>Bn})\right\}=0.
\label{equ:green2}
\end{equation}
The regular solution for positive $E$ is obtained  as described in the previous section, but taking an outgoing Coulomb wave as the boundary condition. 
On the other hand, the irregular solution cannot be expanded in the basis (\ref{equ:lagrange}), since these Lagrange functions are regular at the origin. We use instead the alternative irregular basis defined in the interval $[a_1,a_2]$,
\begin{equation}
\begin{split}
\tilde\varphi_{i}(r)&=(-)^{N+i} (\Delta ax_{i}(1-x_{i}))^{1/2}
\\&\times \frac{P_{N}((2r-a_{1}-a_{2})/\Delta a)}{r-\Delta a x_{i}-a_{1}},
\end{split}
\end{equation}
where we have dropped the factor $r/ax_i$, and $\Delta a= a_{2} -a_{1}$. The generalization of the R--matrix method to arbitrary intervals with $a_1\neq0$ is described in ~\cite{Descouvemont_RPP2010}, and we take advantage of the fact that the value of the irregular solution $g_l(r)$ is known for small $r$.  We thus avoid to deal with the numerical difficulties associated with the divergent behavior of $g_l(r)$ as $r\to0$. As we pointed out in the previous section, the Green's function can also be computed for bound states by enforcing a Whittaker boundary condition for the regular solution.

\section{Numerical details}
\begin{figure}[t!]
	
	\includegraphics[width=.45\textwidth,height=6.8cm]{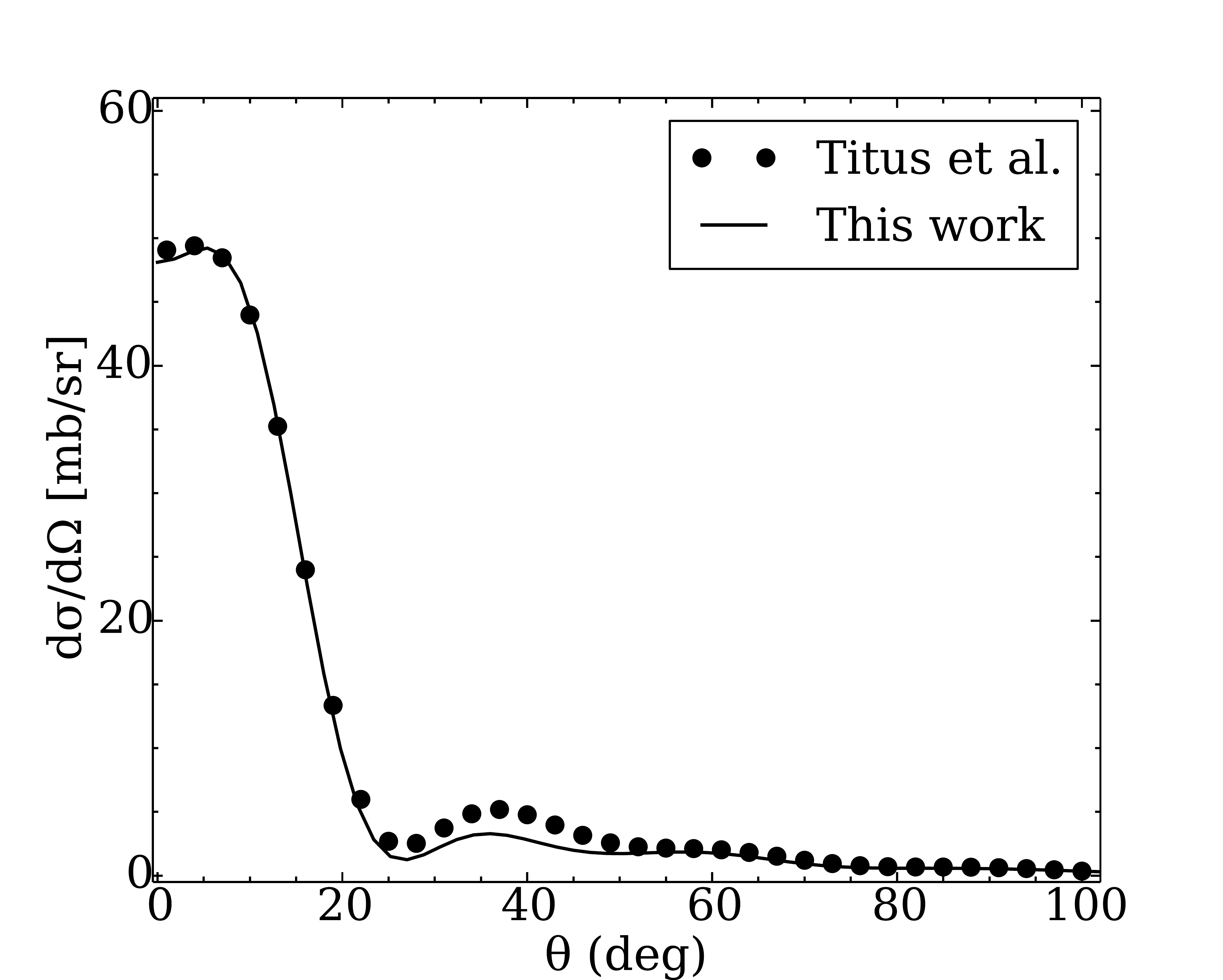}
	\caption{Angular differential cross section corresponding to the $^{48}$Ca($d$,$p$) reaction with an $E_d=20$ MeV deuteron beam. We compare our results (solid line) with the calculation of Titus \textit{et al.} (dots) ~\cite{Titus_CPC2016}. } 
	\label{fig:benchmark} 
\end{figure}

\begin{table*}[ht]
		
		
\begin{tabular}{|l|l|l|l|l|l|l|l|l|l|l|l|}
\hline
 & \multicolumn{10}{c|}{LEP}\\ \hline 
\rule{0pt}{9pt}
		 & \multicolumn{2}{c}{A=$^{16}$O}
	    & \multicolumn{6}{|c|}{A=$^{40}$Ca } & \multicolumn{2}{c|}{A=$^{208}$Pb }   \\ [0.15em] \hline \hline 
	    \midrule	
	  &  $U_{An}$ &  $U_{Ap}$  &  $U_{An}$ & $U_{Ap}$  & $U_{An}$ & $U_{Ap}$  & $U_{An}$&  $U_{Ap}$ & $U_{An}$ &  $U_{Ap}$ \\\hline
	 E$_{n}$/$E_{p}$  (MeV)	 & 2.500 &  14.000  &   3.000 & 7.000  & 3.000 & 15.000  & 10.000 & 40.000  &  3.500 &  14.000 \\\hline
		V$_{v}$ (MeV)  & 41.500   &  43.971 & 	47.441	& 46.040 & 42.902 &	44.311 & 35.100  & 36.881  & 34.451  & 39.100 \\\hline
	r$_{v}$ (fm)   &  1.340  & 1.311  & 1.271  &	1.291  & 1.270  &	1.290  & 1.261  & 1.271  & 1.250 & 1.253 \\\hline
		a$_{v}$ (fm)  & 0.601&  0.620  & 0.671 & 0.62 & 0.640  & 0.619 &  0.632 &  0.616 & 0.601 & 0.615  \\\hline 
	W$_{d}$ (MeV) & 10.001 &  9.891  & 14.501 & 10.201 & 9.132   &	10.051 & 8.030  &  8.741 & 8.291  & 9.070 \\\hline
		r$_{d}$ (fm)  & 1.290	& 1.262  & 1.361 &	1.240 &	1.281  &	1.250 &  1.241 & 1.243  & 1.232  & 1.241   \\\hline
		a$_{d}$ (fm) &	0.401 & 0.440 & 0.361 & 0.446 &	0.441 &	0.431 & 0.442  & 0.425  & 0.431 &  0.422 \\\hline 
		V$_{so}$ (MeV)  &	7.180 & 0.000  & 7.180	 & 0.000	 &	7.180 & 0.000 &  7.180 & 0.000  & 7.180 &  0.000 \\\hline
	r$_{so}$ (fm)  &	 1.220 & \quad---\quad & 1.220 &\quad---\quad &	1.220  & \quad---\quad & 1.220  & \quad---\quad  & 1.220  & \quad---\quad  \\\hline
		a$_{so}$ (fm)  & 0.650  &  \quad---\quad & 	0.650 & \quad---\quad & 0.650 & \quad---\quad & 0.650  & \quad---\quad & 0.650 & \quad---\quad  \\\hline
	r$_{c}$ (fm)   & 1.220  & 1.220  & 	1.220  &	1.220 & 	1.220 &	1.220 & 1.220  & 1.220   & 1.220 & 1.220  \\\hline
	   				 		
		\end{tabular}
		\caption{Local equivalent potential (LEP)  parameters used in this work, according to the parametrization given in Eq.~(\ref{equ:wood-saxson}). The neutron and proton energies at which the local equivalent fitting procedure has been made are indicated in the corresponding column headers. The three $^{40}$Ca column pairs refer to the parameters used in the $E_{d}$=10 MeV, $E_{d}$=20 MeV and $E_{d}$=50 MeV calculations, respectively from left to right.}
			\label{tab:table_parameters}	
		\end{table*}
We performed a systematic study for $(d,p)$ on $^{16}$O, $^{40}$Ca, $^{48}$Ca and $^{208}$Pb at various beam energies, $E_{d}$=10 MeV, 20 MeV and 50 MeV. The optical potential parameterization for $U_{An}$ and $U_{Ap}$ used in final proton and the neutron channel is the non--local Perey-Buck~\cite{Perey_NP1962}.  For the deuteron optical potential we have used a global phenomenological parametrization \cite{Han:06}.

As in~\cite{Titus_prc2014}, in order to assess
the effects of nonlocality, we compare those results with the calculations obtained using local equivalent potentials (LEP) for $U_{An}$ and $U_{Ap}$. For each $(d,p)$ reaction studied (characterized by the target $A$ and the deuteron beam energy $E_{d}$), we estimate the neutron and proton energies corresponding to the the peak of the energy distribution of the NEB cross section.  For those specific energies, we use {\sc sfresco}~\cite{fresco} to fit the elastic scattering generated with the Perey-Buck potential, with a local interaction including a real volume Woods-Saxon term and an imaginary surface term (as in~\cite{Titus_prc2014}). In the fitting, we do not fit the spin-orbit and Coulomb terms; these are fixed as in  the Perey-Buck potential. Then, for each $(d,p)$ reaction, we use for the whole range of energies of the final state, the same LEP parameters as those extracted for neutron and proton at the estimated energies as described above.

The local Wood-Saxon potential is described with the standard parametrization,
\begin{equation}
\begin{split}
V^{\text{LEP}}_{\text{WS}}(R)&=-V_{v}f(R,r_{v},a_{v})+\mathrm{i} 4a_{d}W_{d}\frac{d}{dR}f(R,r_{d},a_{d})\\+&V_{so}\frac{1}{R}\frac{d}{dR}f(R,r_{so},a_{so})2\mathbf{l.s}+V_{c}(R)
\end{split}
\label{equ:wood-saxson}
\end{equation}

where
\begin{equation}
f(R,r,a)=\left[  1+ \mathrm{exp} \left( \frac{R-rA^{1/3}}{a} \right)  \right]^{-1}
\end{equation}

and $V_{c}$ is the Coulomb potential generated by a uniformly charged sphere,

\begin{equation}
V_{c}(R)=\begin{cases}\frac{Z_{1}Z_{2}e^{2}}{2}(3-\frac{R^{2}}{R^{c}_{2}}) & \text{if} \quad  R <R_{c}\\ \frac{Z_{1}Z_{2}e^{2}}{R} &  \text{if} \quad  R \geq R_{c} \end{cases} 
\end{equation} where the radius $R_{c}$ is given by $R_{c} = r_{c} A^{1/3}$. Here $V_v, r_v, a_v$ are the parameters for the depth, radius and diffuseness for the real volume term; $W_d, r_d, a_d$ are the depth, radius and diffuseness for the imaginary surface term, $V_{so}, r_{so}, a_{so}$ are the depth, radius and diffuseness for the spin-orbit term, and $r_c$ is the Coulomb radius. The resulting LEP parameters can be found in Table~\ref{tab:table_parameters}.

Next, we define the model space needed for converged results. The nucleon scattering waves and the neutron's Green's function ($\chi_p$, $\Phi_n$, $G_B^{\text{opt}}$) are obtained including $N=60$ Lagrange basis functions in the R--matrix expansion, and using as the matching radius $a=30$ fm. For converged NEB cross sections, we need to include partial waves up to $l_{p}=15$ (maximum partial wave number) for the $p-B$ relative motion and $\ell$=8 (final neutron states) for the $n-A$ relative motion. We restrict $l_{d}$ (the relative angular momentum between deuteron and target) to a maximum value of $l_{d}$=15.

As mentioned before, the effect of non--local interactions in $(d,p)$ cross sections populating neutron bound states was studied in ~\cite{Titus_prc2014}. The present paper is an extension of this previous work to describe the population of neutron states in the continuum, but the formalism can also make predictions for bound states by considering solutions at negative neutron energy and taking $W_{An}$ to zero (see ~\cite{Potel_PRC2015}). We use this fact to perform a benchmark of the new tools developed in the current study against those developed in ~\cite{Titus_prc2014,Titus_CPC2016}. In  Fig.~\ref{fig:benchmark} we show the angular distribution for the population of the ground state of $^{49}$Ca following $(d,p)$ with $E_{d}$=20 MeV. We take the same parameters as ~\cite{Titus_PRC2016}. Our results are shown with the solid line and agree well with those from Titus et al.~\cite{Titus_CPC2016}.  Due to numerical difficulties associated with the iterative method implemented in [28], the corresponding calculation is not fully converged. This is the origin of the small discrepancies found around 40$^\circ$.

\section{Predictions for the NEB cross sections}
\begin{figure*}
	\includegraphics[width=.45\linewidth]{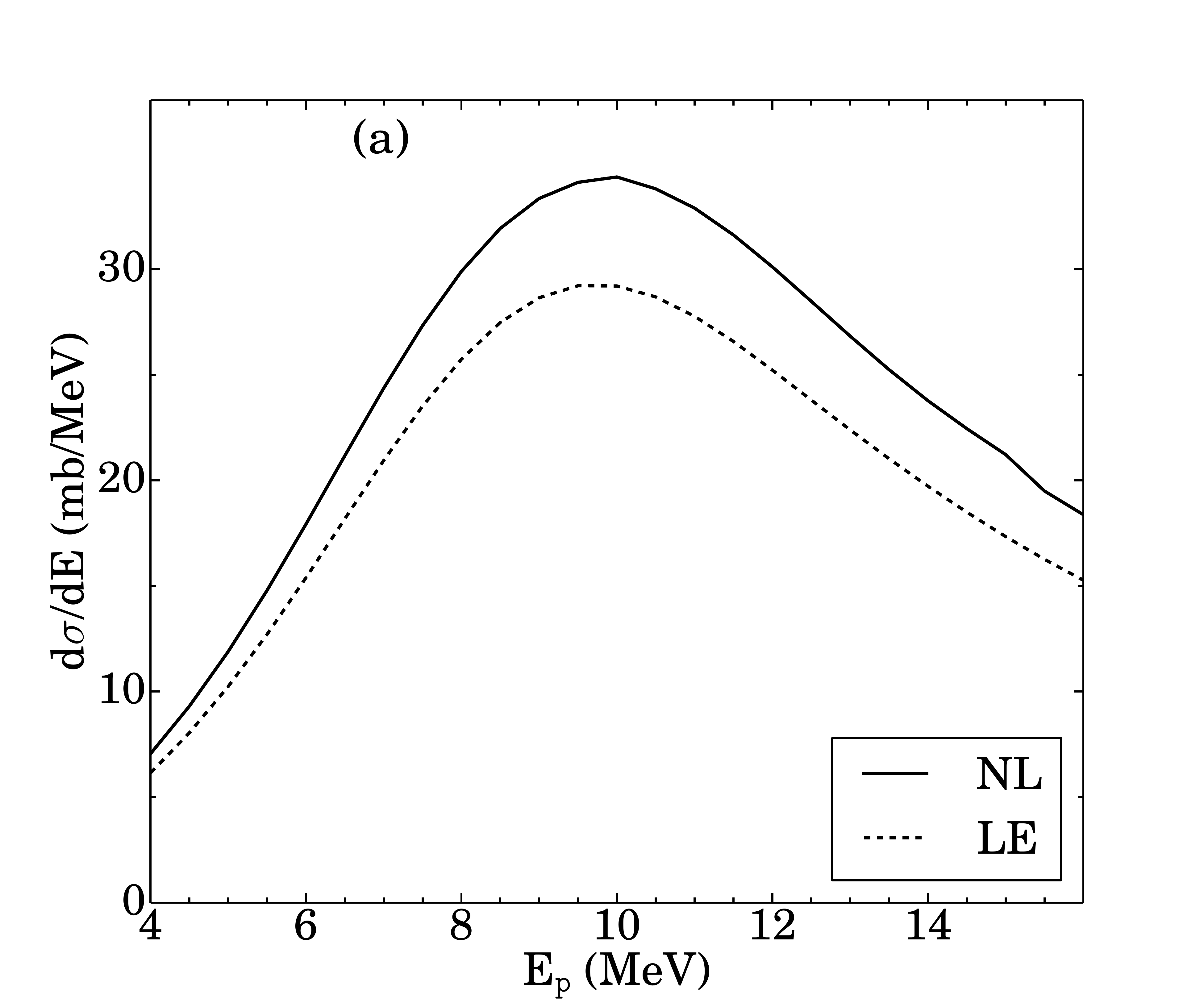}\quad
	\includegraphics[width=.45\linewidth]{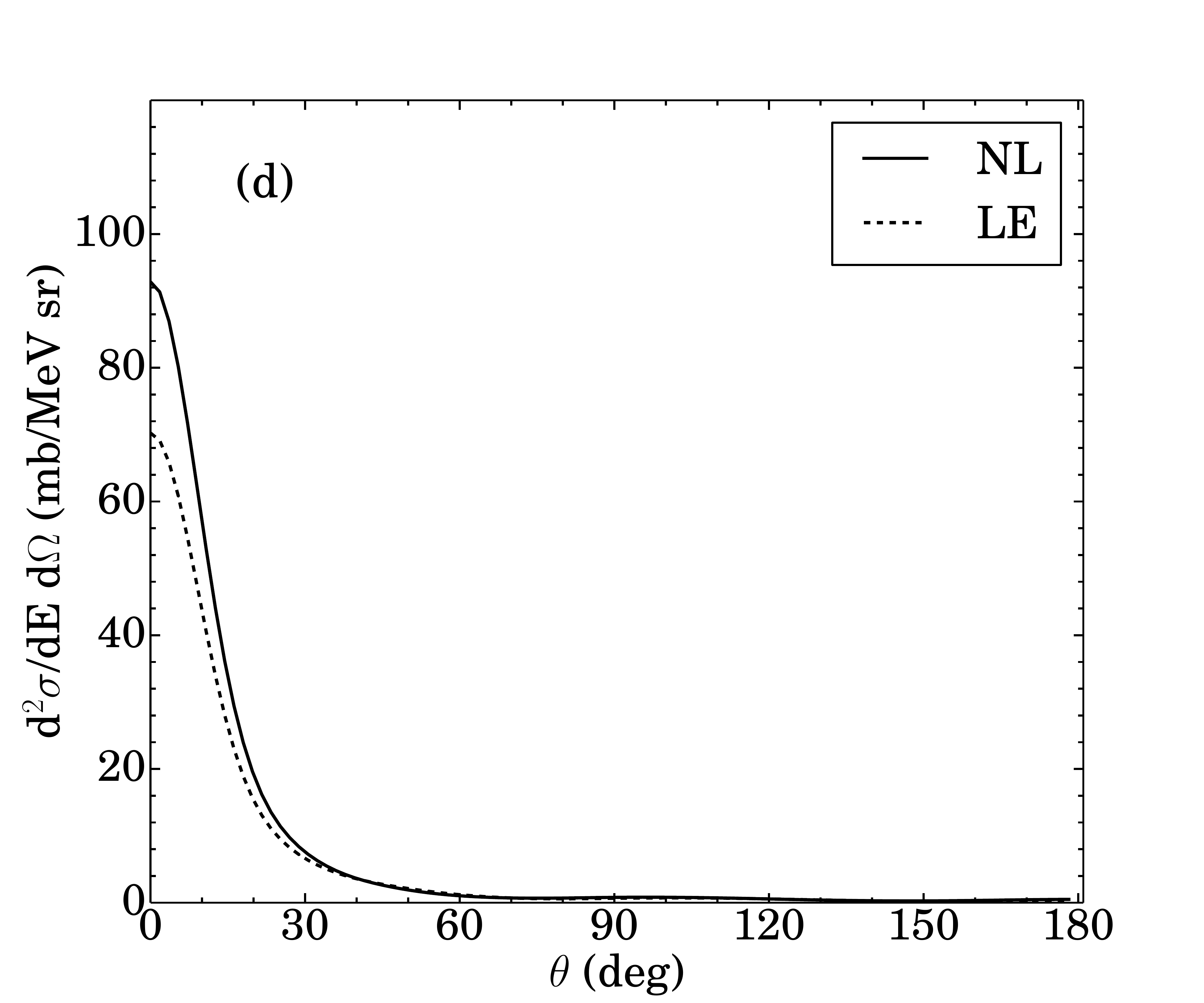}
	\bigskip
	\includegraphics[width=.45\linewidth]{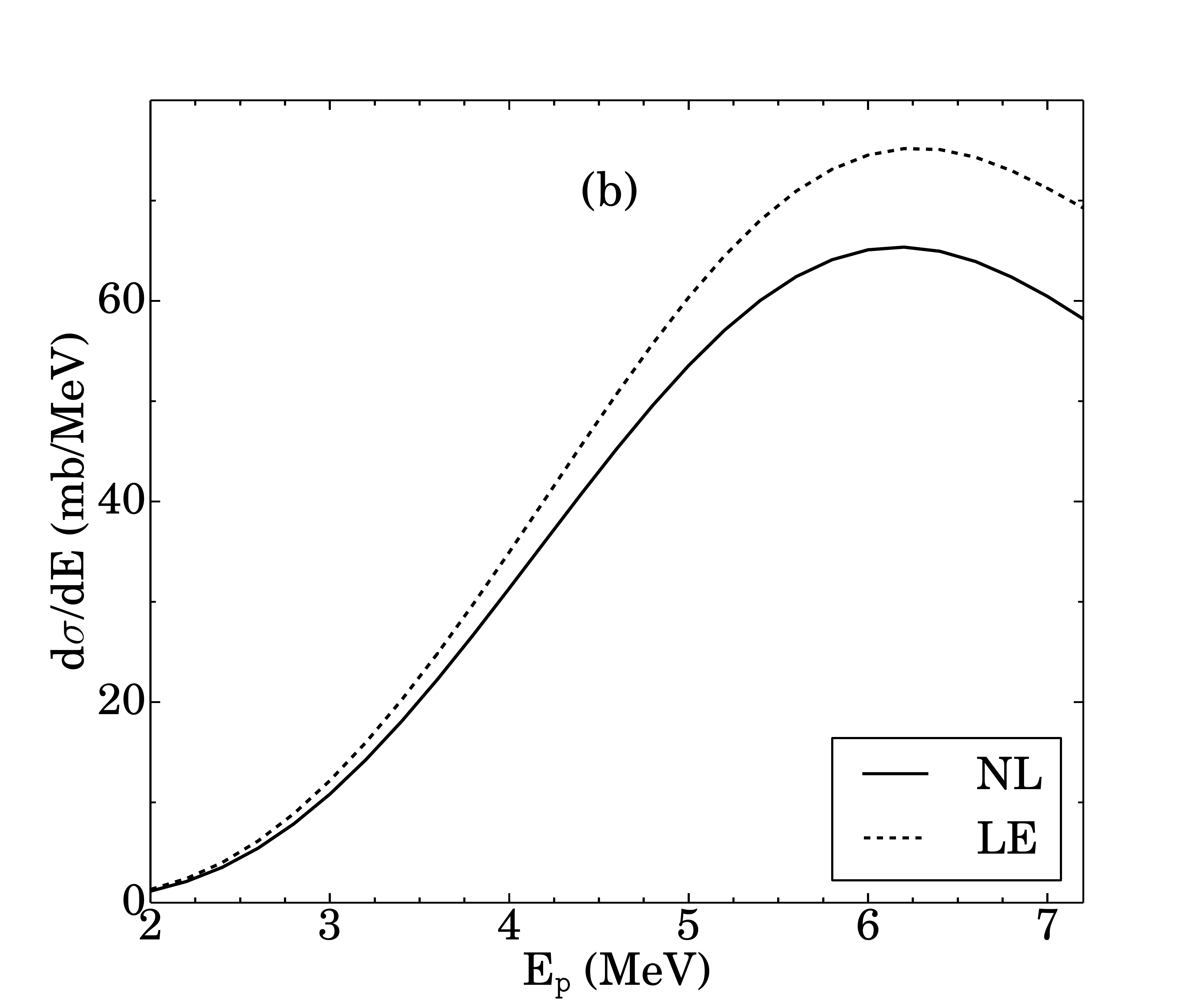}\quad
	\includegraphics[width=.45\linewidth]{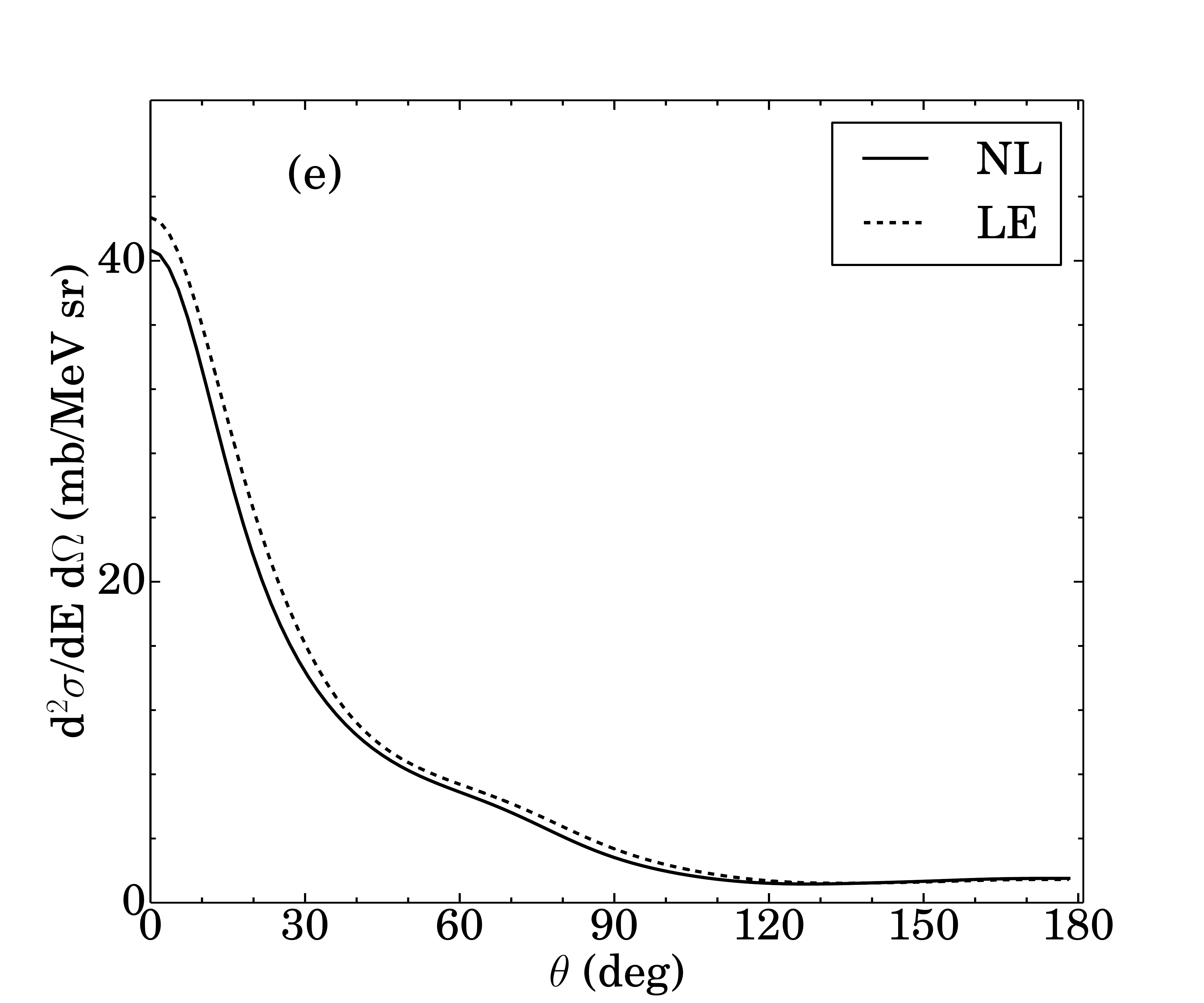}

	\bigskip
	
	\includegraphics[width=.45\linewidth]{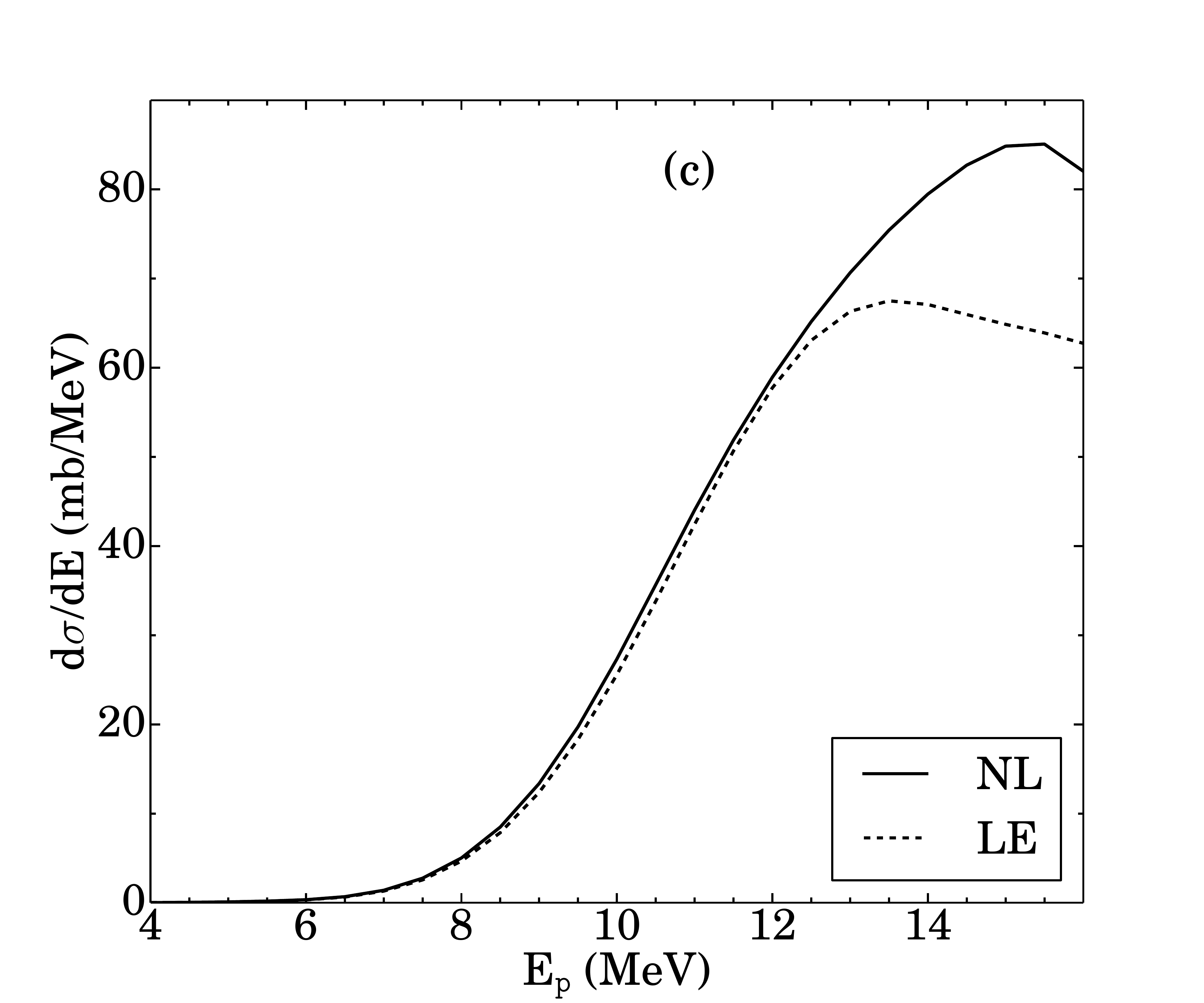}\quad
	\includegraphics[width=.45\linewidth]{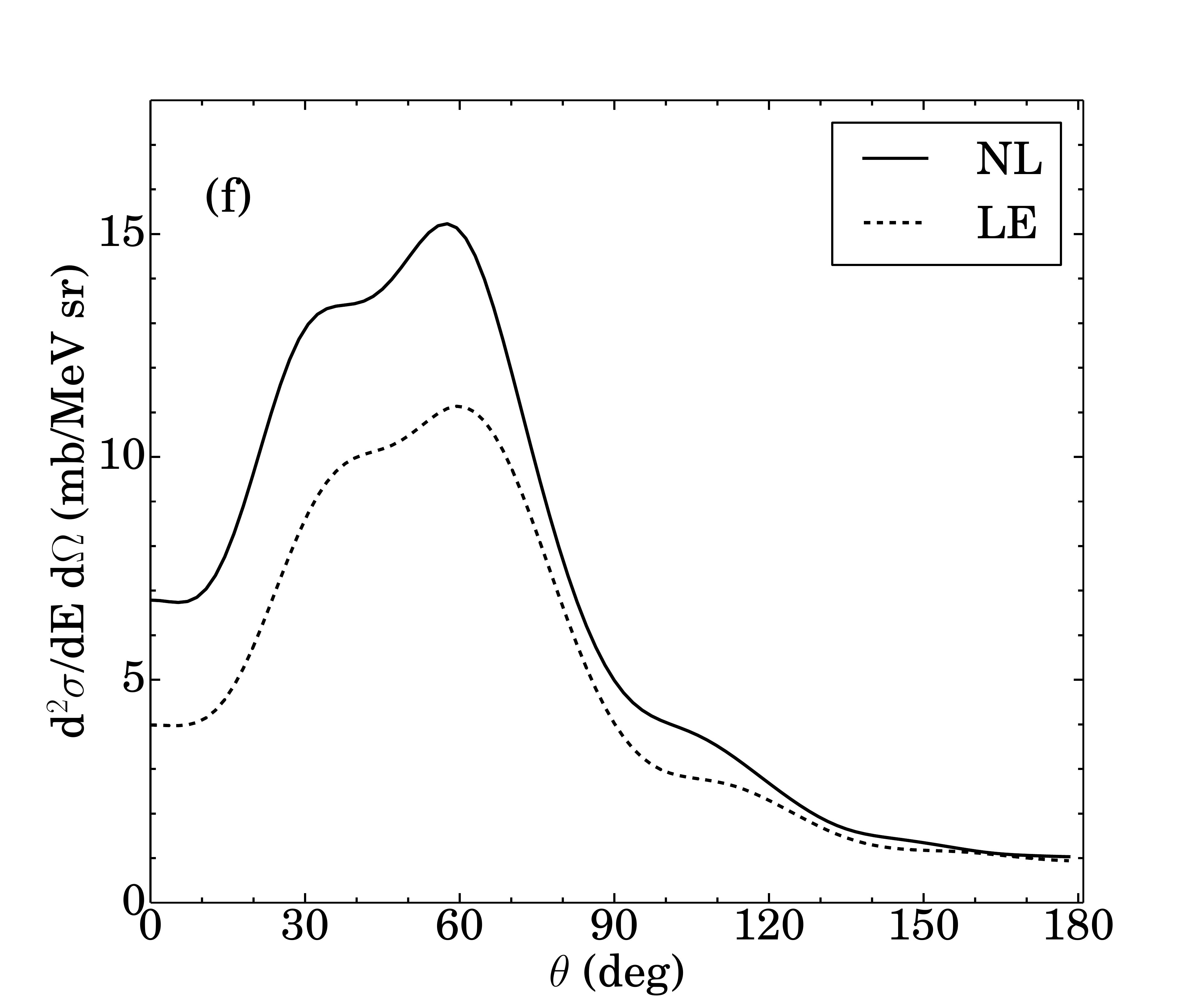}
	\caption{We present in this figure an overview of the differences between the local (LE, dashed line) and non--local (NL, solid line) calculations. In the left column we show the energy differential cross section of the detected proton corresponding to NEB, as a function of the proton energy, for (a): $^{16}$O$(d,p)$ at $E_d=$20 MeV, (b): $^{40}$Ca$(d,p)$ at $E_d=$10 MeV and (c): $^{208}$Pb$(d,p)$ at $E_d=$20 MeV. In the right column, we show the double differential cross section  of the detected proton corresponding to NEB as a function of the center of mass angle for the same reactions  (d): $^{16}$O$(d,p)$ at $E_d=$20 MeV, (e): $^{40}$Ca$(d,p)$ at $E_d=$10 MeV, and (f): $^{208}$Pb$(d,p)$ at $E_d=$20 MeV.  These last three calculations were performed at $E_p$=10.5 MeV for LE and 10.5 MeV for NL (d); $E_p$=5.9 MeV for LE and 6.2 MeV for NL (e); $E_p$=13.6 MeV for LE and 14.7 MeV for NL (f).}
	
	\label{fig:example_cross_section} 
\end{figure*}

Given the strong interest in $(d,p)$ for surrogates, and the parallel between the NEB cross section and the transfer cross section to bound states \cite{Titus_prc2014}, here we focus on the NEB contributions to the $(d,p)$ cross section. To illustrate the overall effect, we first show in Fig. \ref{fig:example_cross_section} the energy and angular distributions for $^{16}$O$(d,p)$ at $E_{d}$=20 MeV (panels a and  d), $^{40}$Ca$(d,p)$ at $E_{d}$=10 MeV (panels b and e), and $^{208}$Pb$(d,p)$ at $E_{d}$=20 MeV (panels c and f).
The plots show the predictions using the non--local interactions, labeled NL (solid lines), and those using the LEP, labeled LE (dotted lines). For all cases, the effects of nonlocality are significant. This overall conclusion is consistent with the studies for bound states ~\cite{Titus_prc2014}. However, given the complexity of the reaction theory, understanding the source for these large effects will prove to be more challenging than in \cite{Titus_prc2014}.

\begin{figure*}
	\centering
	
	\includegraphics[width=.45\linewidth]{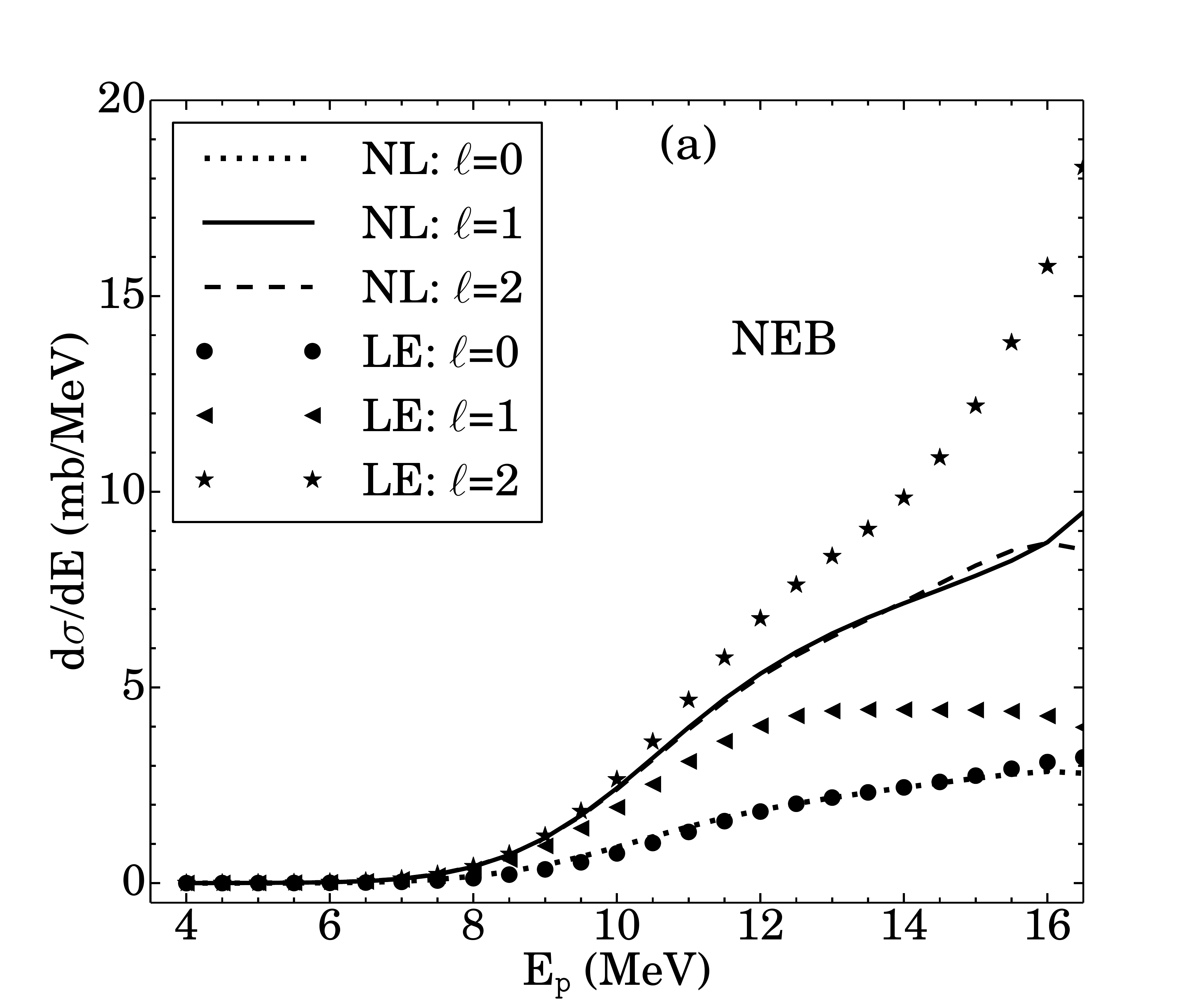}\quad
	\includegraphics[width=.45\linewidth]{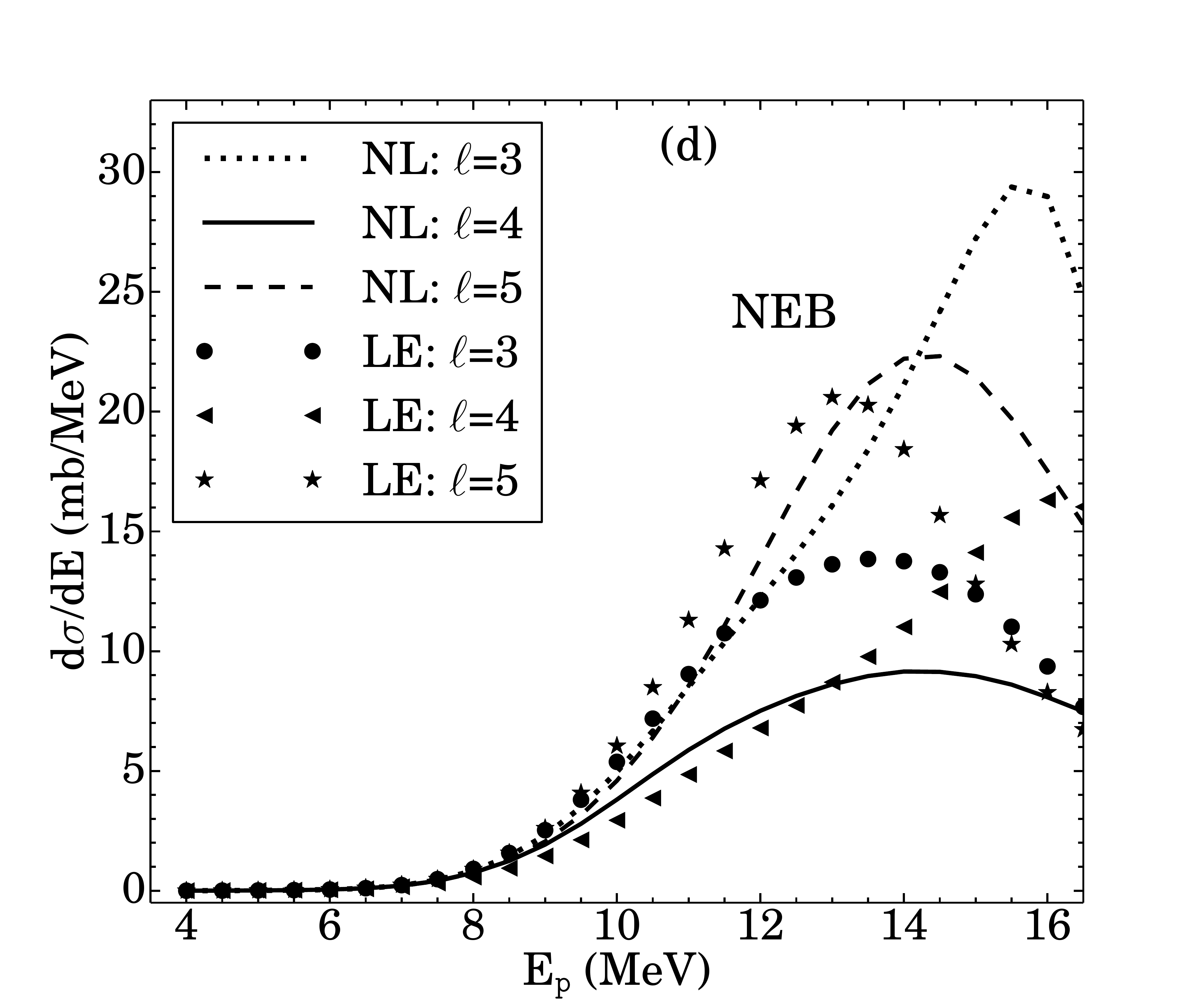}
	
	\bigskip
	
	\includegraphics[width=.45\linewidth]{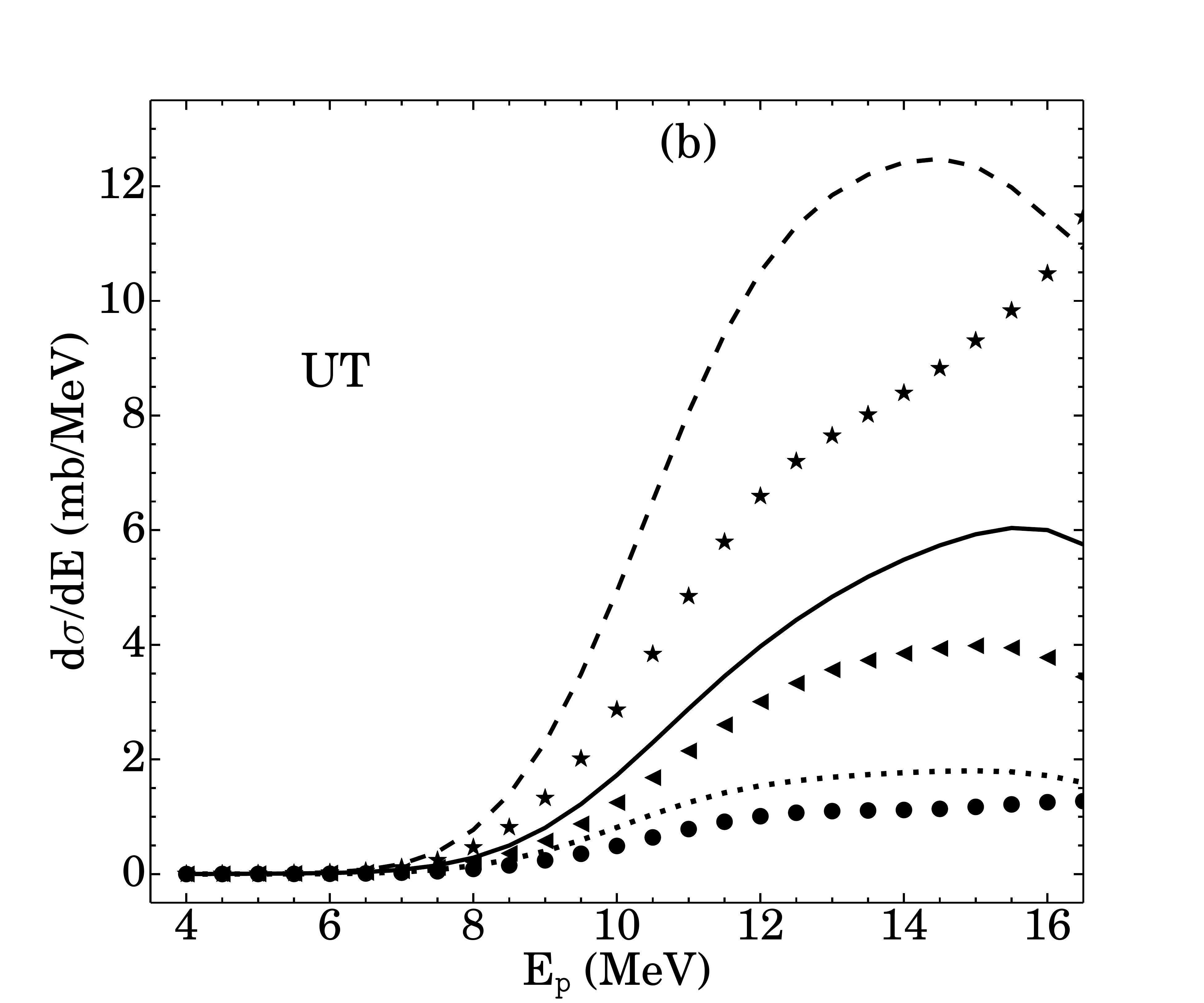}\quad
	\includegraphics[width=.45\linewidth]{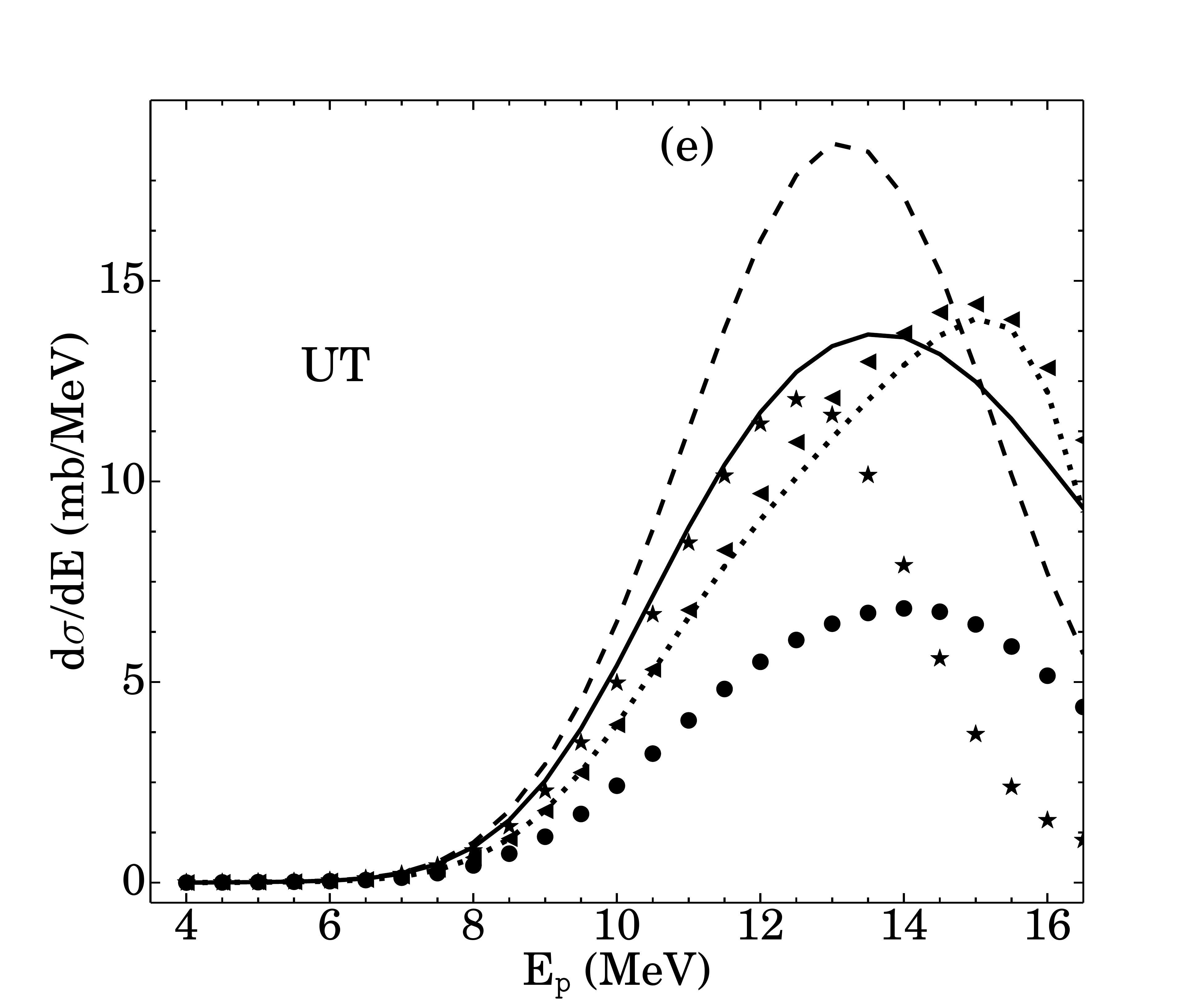}

	\bigskip
	
	\includegraphics[width=.45\linewidth]{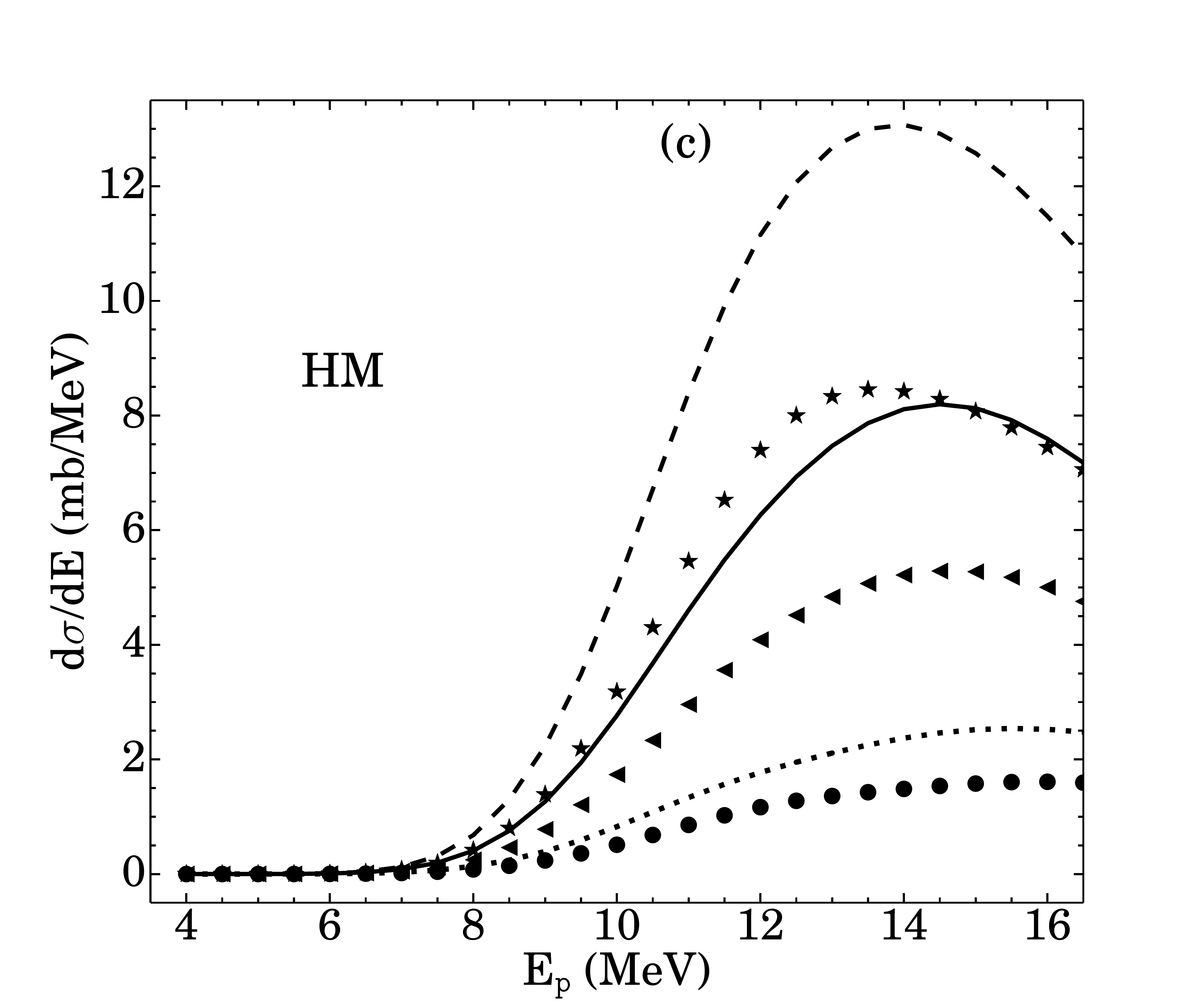}\quad
	\includegraphics[width=.45\linewidth]{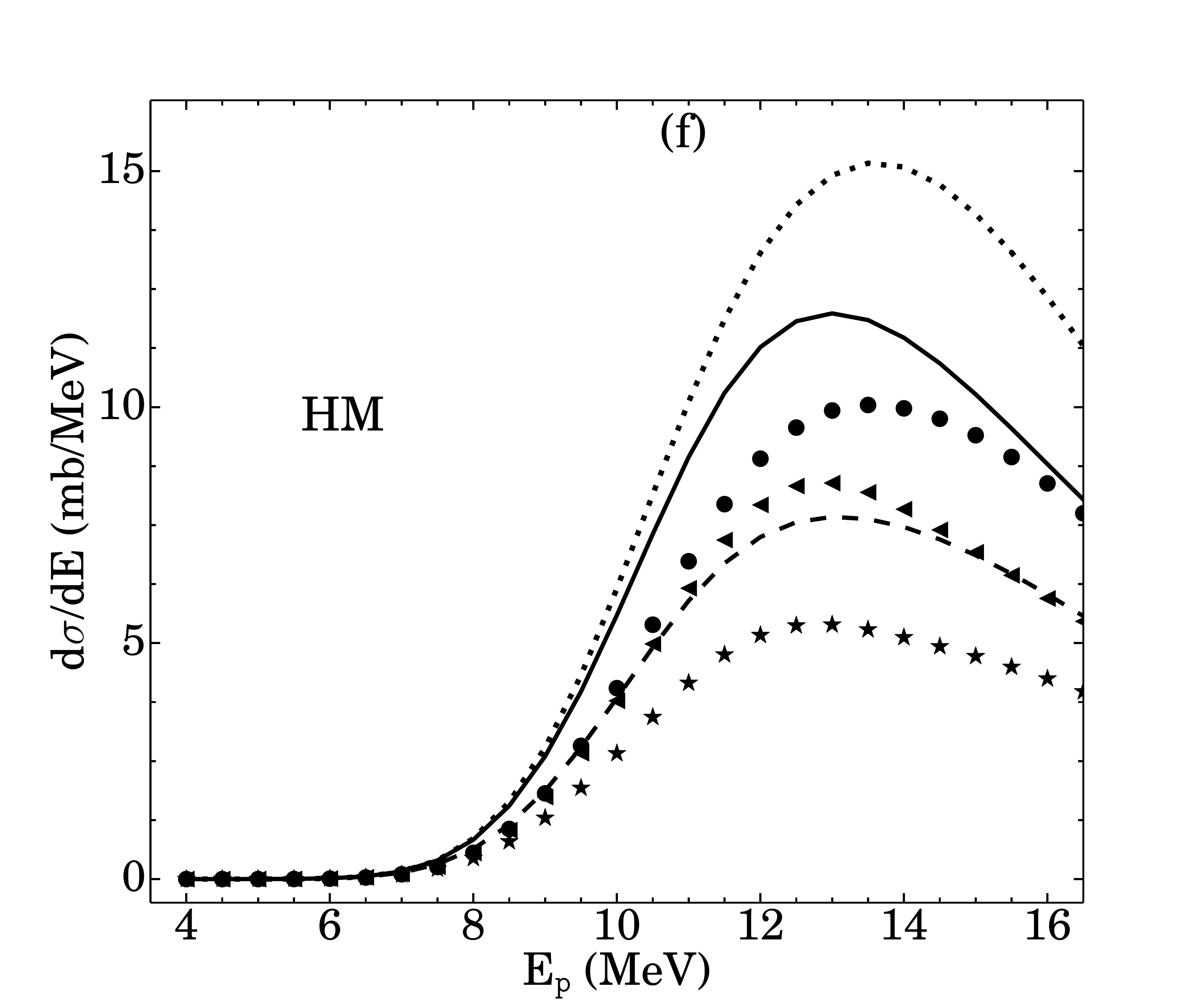}
	\caption{Energy differential cross section of the detected proton as a function of the proton energy, for $^{208}$Pb$(d,p)$ at $E_d=20$ MeV, corresponding to the local equivalent (LE, symbols) and non--local (NL, lines) calculations. In the first row (panels (a) and (d)) we show the total NEB cross section (see Eq. (\ref{equ:cross_section})). The second row (panels (b) and (e)) corresponds to the UT term (see Eq. (\ref{equ:cross_section_UT})), while in the third row  (panels (c) and (f)) the contribution of the non--orthogonality term HM (see Eq. (\ref{equ:cross_section_HM})) is shown. The left column corresponds to the contributions of neutron--target orbital angular momenta ranging from $\ell$=0 to $\ell$=2. In the right column we show the contributions of $\ell=3,4,5$.}
	\label{fig:total_cross_section} 
\end{figure*}

We next consider the contributions to the NEB, UT and HM terms from different partial waves (neutron--target angular momentum $\ell$). We remind here that  in addition to the UT and HM term, there is also an interference term (see Eq. (\ref{equ:cross_section})). In Fig.~\ref{fig:total_cross_section}, and for the $^{208}$Pb $(d,p)$, $E_{d}$=20 MeV reaction, we present the energy distributions for each $n-A$ partial wave  ($\ell=0-5$) when using non--local interactions, and compare them with the corresponding results using LEPs. While for some partial waves, nonlocality has a negligible effect, for others the effect is very large and  changes the shape of the distribution. Table~\ref{tab:table1} contains the partial wave contributions of the  NEB cross section for $^{40}$Ca$(d,p)$ for $E_{d}$=50 MeV at the peak of the energy distribution.  We show results obtained with the LEPs (first row), followed by the percentage difference relative to the local, for the NEB cross sections at the peak of the energy distribution in three different situations: nonlocality is only included in $U_{Ap}$ (NL$_{\text{p}}$, second row), nonlocality is only included in $U_{An}$ (NL$_{\text{n}}$, third row) and nonlocality is included in both (NL, forth row). The proton energy at which these percentage differences are computed is also provided in the second column of Table~\ref{tab:table1}.
We see that for all partial waves, nonlocality in the proton channel reduces the cross section, while nonlocality in the neutron interaction can increase or decrease the cross sections. The competition between these two effects can result in either an increase or a decrease of the cross section.

\begin{table*}[ht!]	
	\begin{tabular}{|l|l|l|l|l|l|l|l|l|l|l|}
		\hline
	\rule{0pt}{9pt}	$\frac{d\sigma}{dE}\rceil^{\text{NEB}}$&$E_{p}$ (MeV)& $\ell$=0 & $\ell$=1 &  $\ell$=2 & $\ell$=3 & $\ell$=4 & $\ell$=5 & $\ell$=6 & $\ell$=7 & total (mb/MeV) \\ [0.15em] \hline 
		LE (mb/MeV) &  26.1 & 0.36 & 1.13  & 2.16  &  2.70  &  5.03 &  5.49 &  2.06 &  0.65 &  19.58 \\
		NL$_{\text{p}}$ ($\%$) & 26.8 & -11.11 & -14.20 &-10.70 & -10.00 & -8.80 & -12.60 &   -15.53   & -16.92  &  -11.54   \\ 
		NL$_{\text{n}}$ ($\%$) & 27.2 & 30.56 & 6.20 & 31.02 & 24.8 & -3.20 &   23.90 &  42.72   & 27.69  &  19.05     \\
		NL ($\%$) & 27.4 & 33.33 & 0.02 & 29.63 & 18.15 & -10.14   &    16.76 &  31.55  & 12.31   & 12.21  \\ \hline \hline
	\rule{0pt}{9pt}	$\frac{d\sigma}{dE}\rceil^{\text{UT}}$&$\tiny{E_{p}}$ (MeV) &  $\ell$=0 & $\ell$=1 &  $\ell$=2 & $\ell$=3  & $\ell$=4 & $\ell$=5 & $\ell$=6 & $\ell$=7 & total (mb/MeV) \\ [0.15em] \hline 
		LE (mb/MeV)  & 26.1  &  0.18  &  0.46  &  0.96  &  1.55   &  1.62  &  1.55   & 0.55 & 0.09 & 6.96 \\		
		NL$_{\text{p}}$ ($\%$) & 26.8 & -22.20 & -23.90 & -21.90 & -17.40 & -23.50 & -32.90 & -42.27 & -44.44  & -26.15 \\ 
		NL$_{\text{n}}$ ($\%$) & 27.2  & -2.20 & -2.20	& -12.50	& 27.10 & 21.60 & -0.70 & 0.00 & -3.33 &  8.95\\ 
		NL ($\%$) & 27.4 & 0.01 & -4.35  & 4.17 &  74.19 &  50.01 &	-6.45 & -25.45 & -22.22 & 24.71 \\ \hline \hline
	\rule{0pt}{9pt}	$\frac{d\sigma}{dE}\rceil^{\text{HM}}  $&$E_{p}$ (MeV) & $\ell$=0 & $\ell$=1 &  $\ell$=2 & $\ell$=3  & $\ell$=4  & $\ell$=5 & $\ell$=6 & $\ell$=7 &  total (mb/MeV) \\ [0.15em] \hline 		
		LE (mb/MeV)  &  26.1 &  0.24  & 0.73 &  1.61 &  3.06  &  3.37 & 3.18 &  2.12  &  0.84  &  15.15    \\
		NL$_{\text{p}}$ ($\%$) & 26.8 & 2.51 &-6.85 & 4.35 &-3.92 &-2.37 &	-2.20 & -14.62	 &  -23.81 & -4.98 \\	
		NL$_{\text{n}}$ ($\%$) & 27.2  &58.99 &	56.16 &	68.94 &	59.15 &	51.04	& 43.71 & 32.08 & 20.24  &  49.05 \\ 
		NL ($\%$) & 27.4 & 58.33 & 42.47	& 20.19 &	50.00 &	47.18 &	39.94 & 15.09	& 0.01  &  41.52 \\ \hline				
		\end{tabular}	
		\caption{Percent differences, with respect to the  LE calculation, of the $^{40}$Ca$(d,p)$ reaction at $E_d=50$ MeV calculated with a non--local proton--target potential (NL$_{\text{p}}$), a non--local neutron--target potential (NL$_{\text{n}}$), and with both  proton and neutron non--local potentials (NL). The upper horizontal block corresponds to the NEB cross section, the second one corresponds to the UT term, and the third one shows the HM contribution. Each calculation has been performed at the proton energy indicated in the second column. We indicate separately the contributions of the different neutron--target orbital angular momenta (columns 3 to 10).}
		\label{tab:table1}
		\end{table*}
		
		\begin{table*}	[ht]
		\begin{tabular}{|l|l|l|l|l|l|l|l|l|l|l|l|}
		\hline
	\rule{0pt}{9pt}	$\frac{d\sigma}{dE}\rceil^{\text{NEB}}$&$E_{p}$ (MeV)& $\ell$=0 & $\ell$=1 &  $\ell$=2 & $\ell$=3 & $\ell$=4 & $\ell$=5 & $\ell$=6 & $\ell$=7 & $\ell$=8&  total (mb/MeV) \\ [0.15em] \hline
		LE (mb/MeV) & 13.6  &   2.34  &  4.44  &  9.19 &  13.86  & 10.00   &  20.01 &  3.64  & 2.62 & 	1.37 &	67.47 \\\hline
		NL$_{\text{p}}$ ($\%$)& 15.0 &  16.67	& -0.90	&   31.96	& -10.46 &	41.00	& -34.98&  -21.43 &	-16.03& -19.71 &-4.13 \\\hline 
		NL$_{\text{n}}$ ($\%$)& 15.5 & 19.23	& 94.37 &-5.88	& 111.98 &	-13.70	& -1.50	&   0.55 &	-4.58 &	25.55 &	26.47 \\\hline 
		NL ($\%$) & 14.7 &  13.68 & 83.33 & -13.06 & 85.43	& -9.00	& 11.44 &  25.27 & 3.05  &	31.39 &	25.92  \\\hline  \hline
	\rule{0pt}{9pt}	$\frac{d\sigma}{dE}\rceil^{\text{UT}}$&$\tiny{E_{p}}$ (MeV) &  $\ell$=0 & $\ell$=1 &  $\ell$=2 & $\ell$=3  & $\ell$=4 & $\ell$=5 &  $\ell$=6 & $\ell$=7 & $\ell$=8&  total (mb/MeV) \\ [0.15em] \hline  
		LE (mb/MeV)  & 13.6  &   1.11  &   3.76 &  8.09  &  6.76  &  13.15  &  9.75   & 1.36 & 1.15 & 0.33 &	45.46 \\\hline 		
		NL$_{\text{p}}$ ($\%$) & 15.0 &  1.80 &	4.26 &	13.10 &	 -6.36	& 9.66	& -62.46 &  -2.21 &	-21.74 &-27.27& -9.63 \\\hline 
		NL$_{\text{n}}$ ($\%$)& 15.5 & 49.55 &	43.62 &	42.03 &	96.75	&  -17.57	 & -1.33  & -44.12 & 36.52 & 121.21&	21.73 \\\hline 
		NL($\%$) & 14.7 & 63.64& 54.79 &	53.89	& 105.30 & -1.75	& 46.77	& -32.35	 & 55.65 &	127.27 &	42.15 
		\\\hline \hline	
	\rule{0pt}{9pt} 
	$\frac{d\sigma}{dE}\rceil^{\text{HM}}  $  &$E_{p}$ (MeV) & $\ell$=0 & $\ell$=1 &  $\ell$=2 & $\ell$=3  & $\ell$=4  & $\ell$=5 &  $\ell$=6 & $\ell$=7 & $\ell$=8&  total (mb/MeV) \\ [0.15em]   \hline		
		LE  (mb/MeV)  &  13.6 &  1.44  &  5.11  &  8.46  &   10.05  &  8.14  &     5.26  & 2.88 &	1.30	& 0.66 &	43.30 \\\hline
		NL$_{\text{p}}$ ($\%$) & 15.0 & 11.11 &	2.54 &	-3.78 &	-6.77 &	-14.74	& -9.70 &  -20.49 &	-15.38&	-13.64 &-7.62
		\\\hline 
		NL$_{\text{n}}$ ($\%$)& 15.5 & 75.69 &	56.56 &	42.08 &	33.63 &	16.95	& 23.57	& 5.21 &	10.77 &	15.15&	32.17
		\\\hline 
		NL ($\%$) & 14.7 & 72.92 &	60.27 &	51.30 &	44.18 &	31.20	& 34.41	&  17.01&	21.54 &	21.21&	41.96
		\\\hline 				
		\end{tabular}
		\caption{Same as Table~\ref{tab:table1}, but for $^{208}$Pb$(d,p)$ at $E_d=20.0$ MeV.}
		\label{tab:table2}	
		\end{table*}

Given that there are three terms composing the total NEB cross section, and in each one of these, nonlocality enters differently (Eq.~(\ref{equ:cross_section})),  it is important to dissect further the results if we want to understand the origins of the overall effects shown in Fig.~\ref{fig:example_cross_section}.
In Table \ref{tab:table1} we  present the same partial wave contributions, either for the direct term (UT)  or for the non--orthogonality term (HM). 
For UT, nonlocality in the proton wavefunction produces a strong reduction independently of the partial waves. The inclusion of nonlocality in $U_{An}$ is more complex, and varies considerably from $\sim 10$\% reduction to $\sim 30$\% increase. Most importantly, these effects interfere in such a way that even if both separate NL$_{\text{n}}$ and NL$_{\text{p}}$ show a decrease in the cross section, one can end up with a total effect that increases the cross section.

The HM term is somewhat simpler to analyse. Nonlocality in $U_{Ap}$ is not very strong, the effects are dominated by the difference in the actual interactions since $W_{An}$ is different for the LEP and the non--local Perey and Buck interaction (see also Fig.~\ref{fig:potential}).

Moreover, we should note that the effects of nonlocality depend strongly on the reaction studied. As an illustration of this fact, Table \ref{tab:table2} contains the same information as Table \ref{tab:table1} but now for the reaction $^{208}$Pb$(d,p)$ at 20.0 MeV. The magnitudes and the signs of the percentage differences can change considerably, indicating a complex balance between the various ingredients, and a strongly non-linear effect.

\begin{figure}
		\includegraphics[width=.45\textwidth,height=6.8cm]{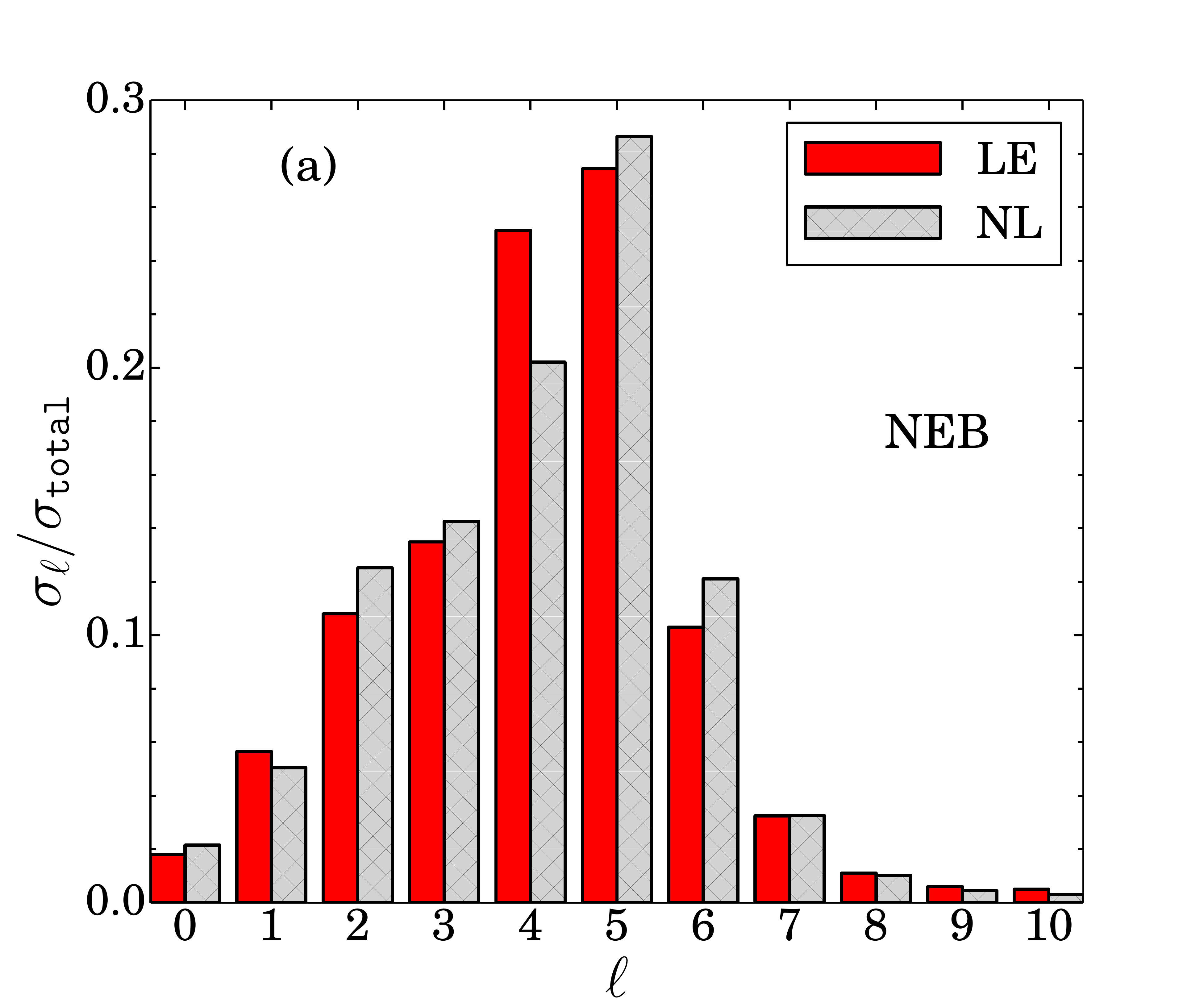}
		\includegraphics[width=.45\textwidth,height=6.8cm]{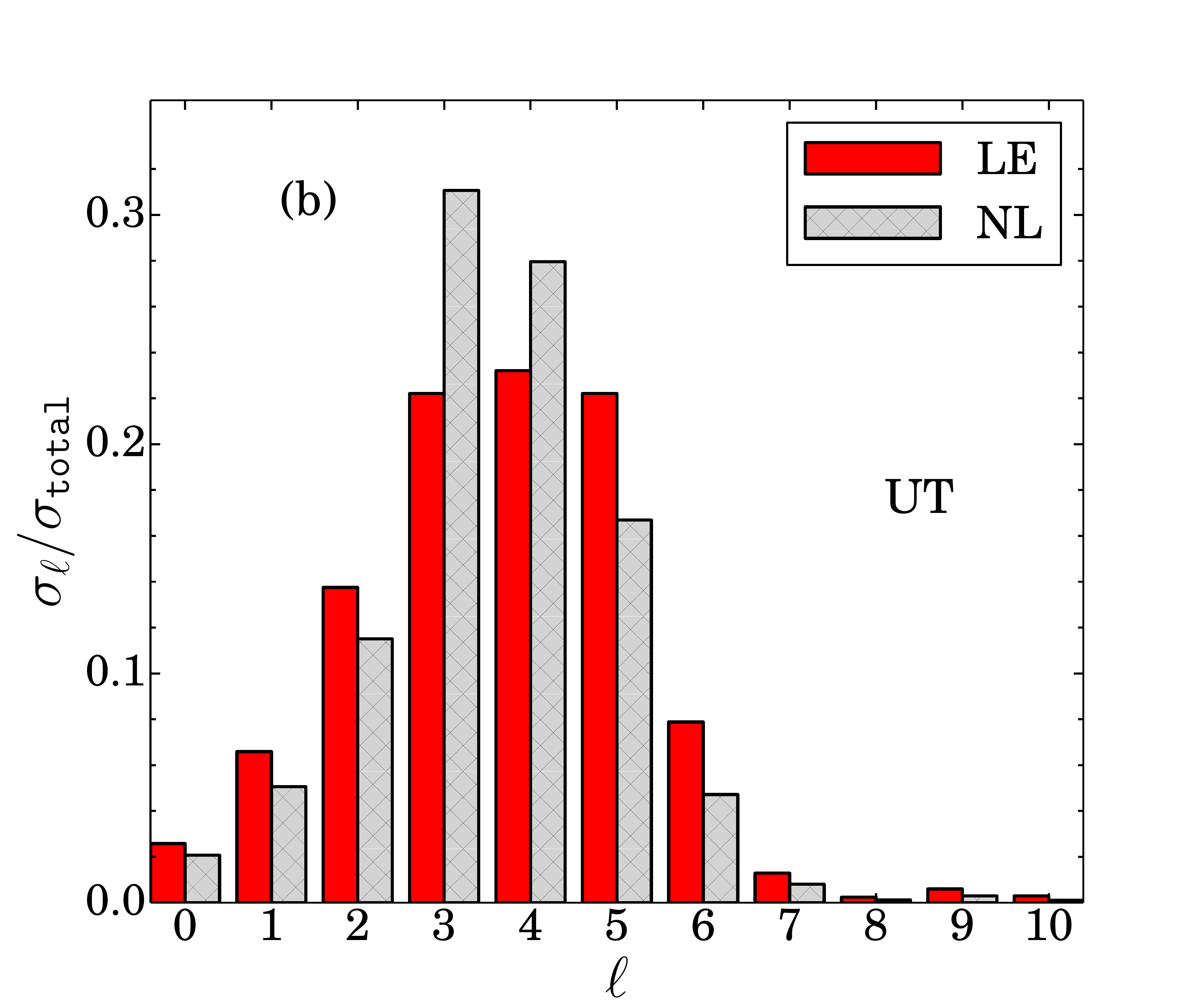}\hfill%
			\includegraphics[width=.45\textwidth,height=6.8cm]{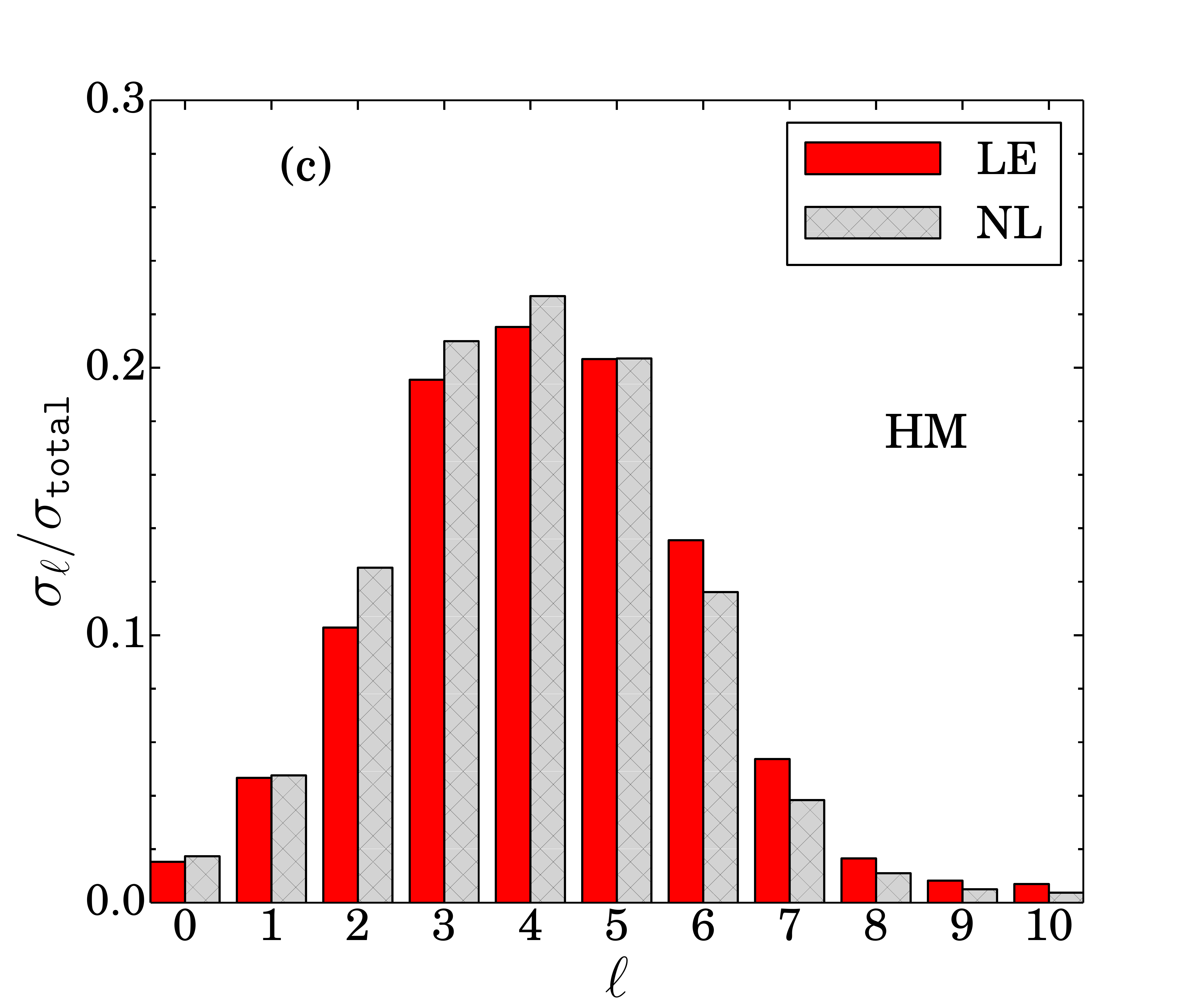}\hfill%
		\caption{Histograms of the relative contributions of the different neutron--target angular momenta to the $^{40}$Ca$ (d,p)$,  $E_d=50$ MeV cross section.  We show the difference between the LE (red, plain) and the NL (grey, hatched) calculation for the NEB (panel (a)), UT (panel (b)) and HM (panel (c)) terms.} 
		\label{fig:percentage}
		\end{figure}

Finally, and keeping in mind that the most important input of this $(d,p)$ reaction theory for the surrogate method is the cross section distribution as a function of the neutron-target partial wave (so-called spin distributions), we show in Fig. \ref{fig:percentage}   the histograms for the ratio of the  cross section corresponding to each neutron partial wave $\ell$ over the total cross section, as a function of $\ell$, for the results obtained with either the LEPs (red, plain) and the non--local interactions (grey, hatched). 
In Fig.~\ref{fig:percentage}a, we show the  NEB cross section while in Fig.~\ref{fig:percentage}b (c) we show the UT (HM) contribution.
The spin distributions shown in Fig. \ref{fig:percentage}  refer to the reaction $^{40}$Ca$(d,p)$ at $E_d=$50 MeV.

As we had seen before,  nonlocality has a sizable effect in the overall magnitude of the cross section. However, the relative importance of the different $\ell$ values is mostly unchanged. As a consequence, the angular differential cross section conserves its shape when including non--local interactions as demonstrated by Figs. \ref{fig:example_cross_section} (d), (e), (f).

\section{Understanding the effects of nonlocality}		
		
So far we have shown that nonlocality has large effects on the NEB cross sections, however it is not easy to interpret those effects in terms of specific physical features of the system. To get a better understanding of the effects of nonlocality, we now turn to the essential ingredients that drive the effects we are seeing, namely the sources  and the actual interactions. Because the various terms in the NEB cross section relate directly  to  matrix elements of $W_{An}$,  the relevant region of space  to consider is the surface region.

We first look at the proton vertex functions, which correspond to the source in the equation for calculating the proton distorted waves $\chi^{l_{p}}_{p}$ (see Eq.~(\ref{equ:schrodinger_nonlocal})). The expressions for the local and non--local proton vertex for a specific $l_{p}$ functions are given by:
		\begin{align}\label{vertex}
		\nonumber &\tilde{\mathcal S}_{\text{LE}}(r)=U_{Ap}(r)\chi^{l_{p}}_p(r),\\
		&\tilde{\mathcal S}_{\text{NL}}(r)=\int U_{Ap}(r,r')\chi^{l_{p}}_p(r')r'^2dr'
		\end{align}
		
\begin{figure}
		\includegraphics[width=.45\textwidth,height=6.8cm]{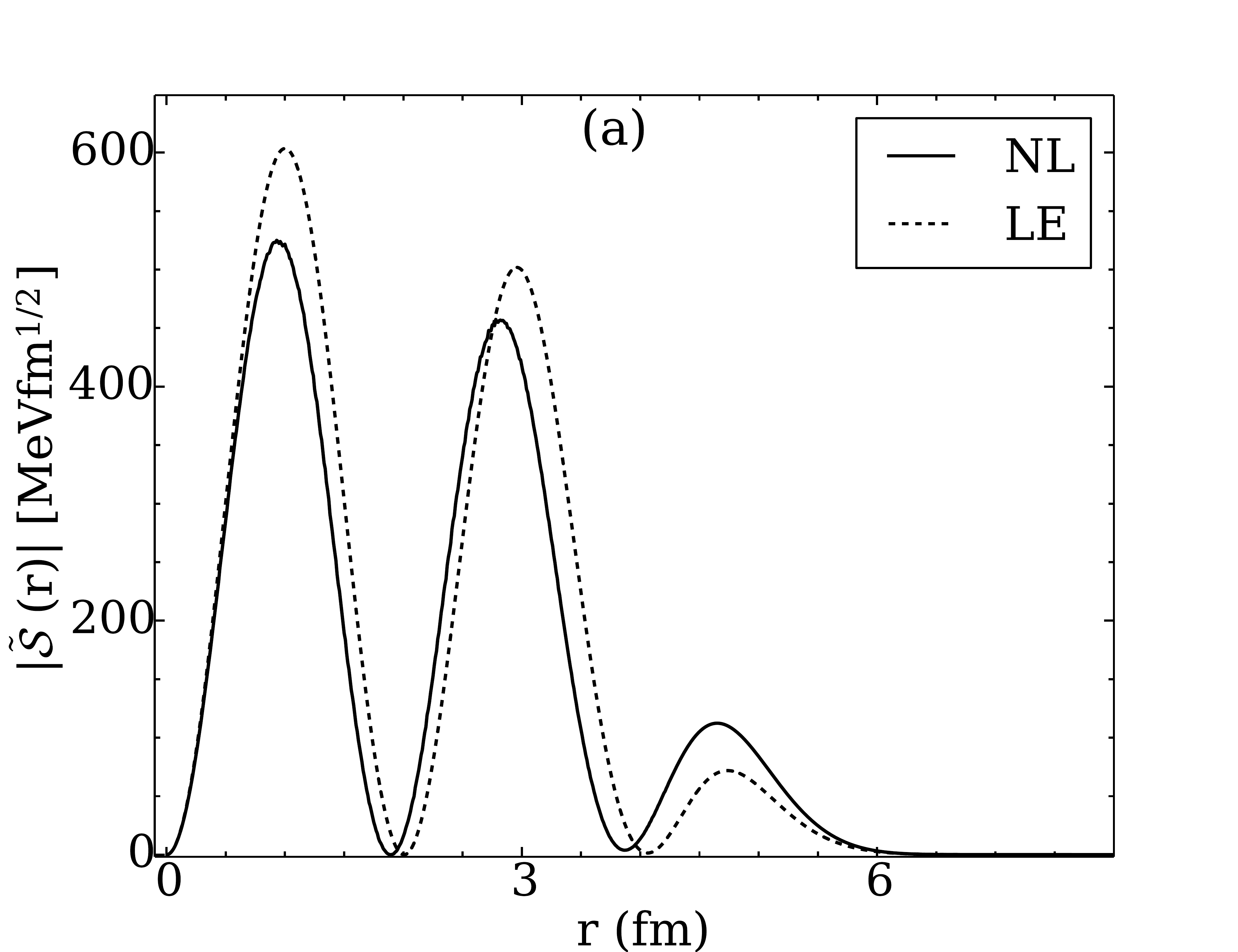}\hfill%
	        \includegraphics[width=.45\textwidth,height=6.8cm]{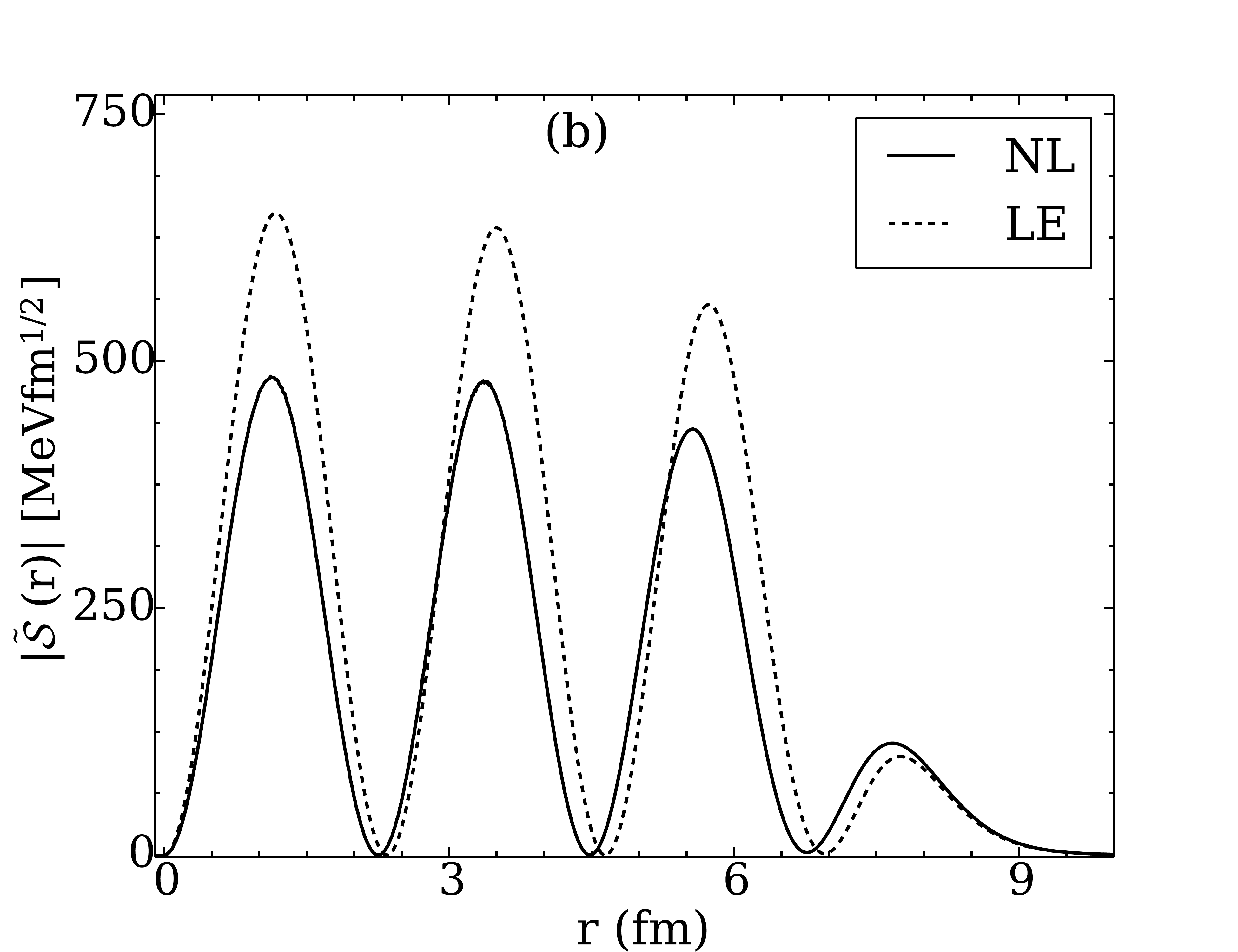}
		\caption{We show in this figure the vertex functions $\tilde{\mathcal S}_{\text{LE}}$ (dashed line) and $\tilde{\mathcal S}_{\text{NL}}$ (solid line) defined in Eq. (\ref{vertex}). Panel (a) corresponds to the $p$+ $^{40}$Ca reaction, with $E_{p}$=26.1 MeV for LE and $E_{p}$=27.4 MeV for NL. In panel (b) we show the results for $p$+$^{208}$Pb, with $E_{p}$=13.6 MeV for LE and $E_{p}$=14.7 MeV for NL. All calculations correspond to the $l_{p}$=0 component of the partial wave decomposition of the vertex functions.}
		\label{fig:proton_integral} 
		\end{figure}

		In Fig.~\ref{fig:proton_integral}, we compare the non--local (solid line) with the local vertex functions defined in Eq. (\ref{vertex}) for $p$+ $^{40}$Ca (panel a) and $p$+$^{208}$Pb (panel b), in both cases for $l_p=0$. Consistently with previous studies, it is observed that the NL vertex function is reduced in the interior region, and enhanced in the exterior, with respect to the local result.
		
\begin{figure}
		\includegraphics[width=.45\textwidth,height=6.8cm]{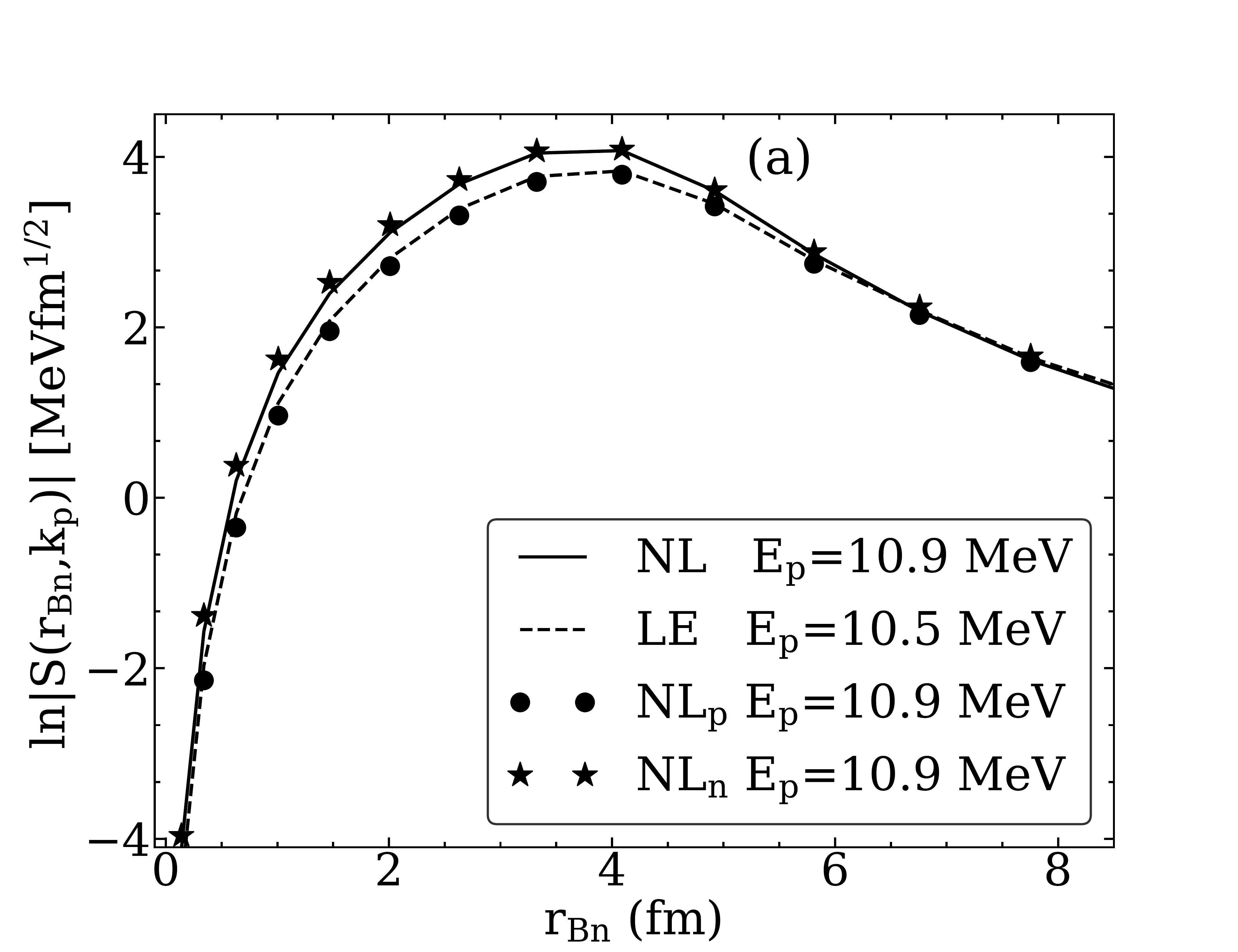}\hfill%
		\includegraphics[width=.45\textwidth,height=6.8cm]{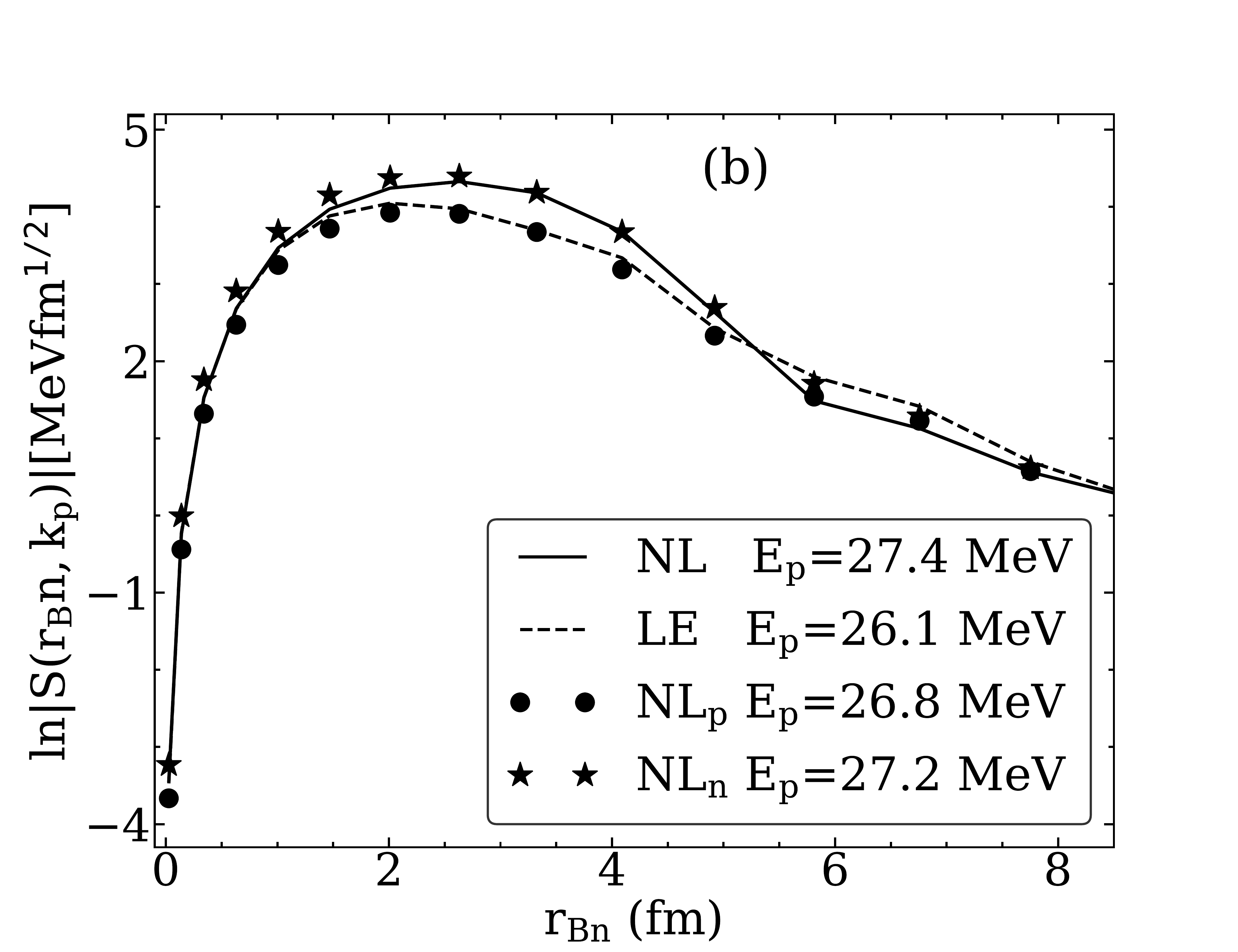}\hfill%
		\caption{Natural logarithm of the absolute value of the source term $S$ (see Eq. (\ref{equ:source})). We show results at specific proton energies (indicated in the figure), for different values of the deuteron energy $E_d$ and of the neutron--target angular momentum $\ell$ corresponding to the partial wave decomposition of $S$. Panel (a): $^{40}$Ca$(d,p)$;  $E_{d}$=20 MeV; $\ell$=3. Panel (b):$^{40}$Ca$(d,p)$; $E_{d}$=50 MeV; $\ell$=2. } 
		\label{fig:source} 
		\end{figure}

We next consider the two elements needed to calculate the neutron wavefunction $\Phi_n$, namely the source $S$ defined in Eq.~(\ref{equ:source}) and the Green's function $G_B^{\text{opt}}$. In Fig.~\ref{fig:source}  we show the  source term $S$  for $^{40}$Ca$(d,p)$ at $E_{d}$=20 MeV (panel a) and  $E_{d}$=50 MeV (panel b). 
Again the results are shown for the case in which we include nonlocality in both the neutron and proton interactions (NL, solid line) and compare to those using the LEPs (LE, dotted line). We also show the  result when including nonlocality only in $U_{Ap}$ (NL$_{\text{p}}$, circles) or  only in $U_{An}$ (NL$_{\text{n}}$, stars). 
The calculations are performed at the proton energy corresponding to the peak of the NEB energy differential cross section, i.e. $E_{p}$=27.4 MeV for NL, $E_{p}$=27.2 MeV for NL$_{\text{n}}$,  $E_{p}$=26.8 MeV for NL$_{\text{p}}$, and  $E_{p}$=26.1 MeV for LE. Nonlocality in the neutron interaction has the effect of increasing the source term in the nuclear interior (comparing LE with NL$_{\text{n}}$), while nonlocality in the proton interaction causes a more modest reduction of the source term in the internal region (comparing LE with NL$_{\text{p}}$). 
		
\begin{figure}
		\includegraphics[width=.45\textwidth,height=6.8cm]{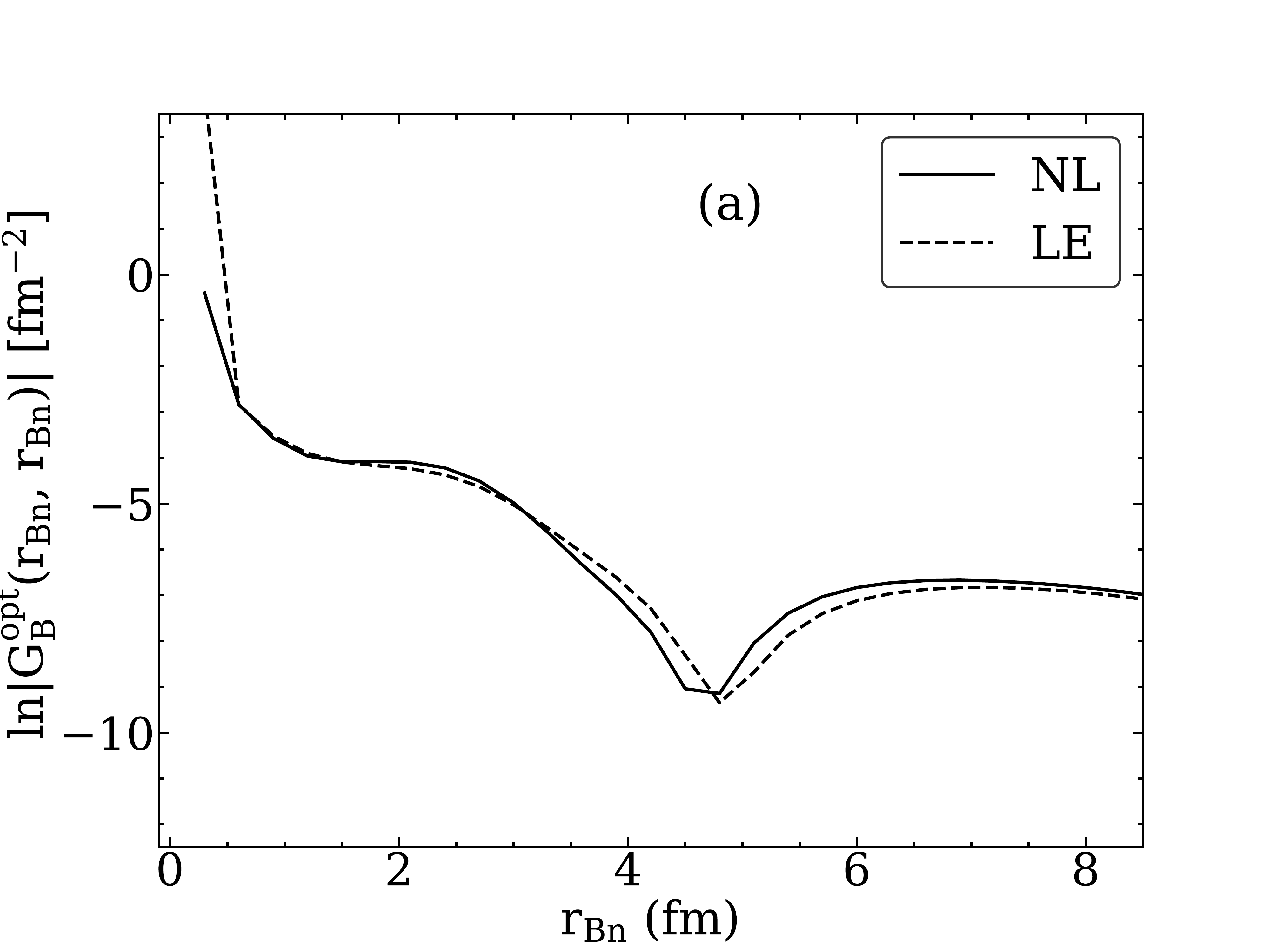}\hfill%
		\includegraphics[width=.45\textwidth,height=6.8cm]{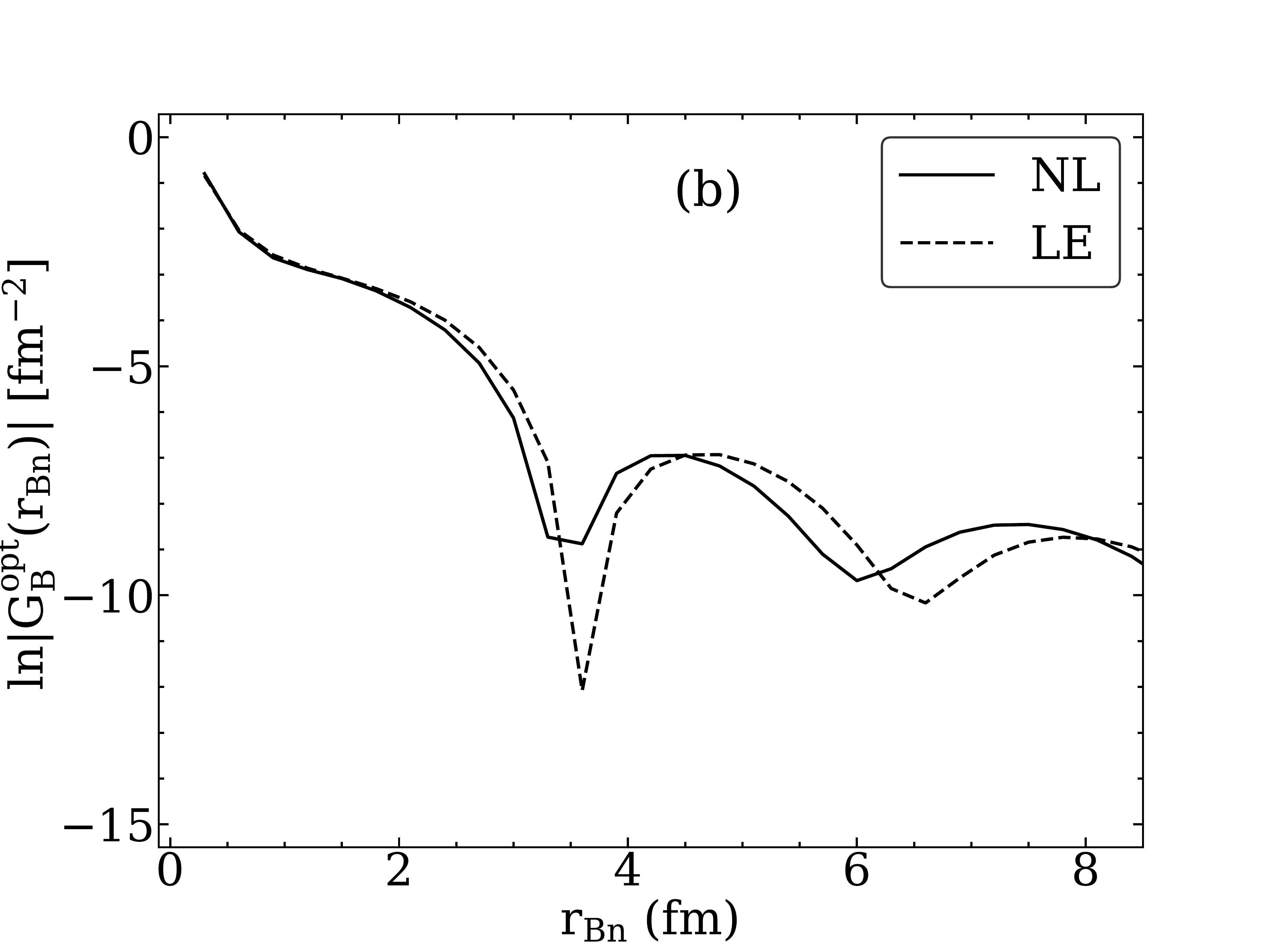}\hfill%
		\caption{Natural logarithm of the absolute value of the diagonal part of the Green's function  as a function of radius for: (a) n+$^{40}$Ca, $E_{n}$=6.74 MeV for LE (6.37 MeV for NL), $\ell$=3 and $j$ =7/2, (b) n+$^{40}$Ca, $E_{n}$=20.43 MeV for LE (19.13 MeV for NL), $\ell$=2,  and $j$=5/2.} 
		\label{fig:green} 
		\end{figure}

In Fig.~\ref{fig:green} we show the diagonal part of the Green's functions $G_{B}^{\text{opt}}(r_{Bn},r_{Bn})$ for $n$ + $^{40}$Ca; $E_{n}$=6.74 MeV for LE (6.37 MeV for NL), $\ell$=3 and $j$ =7/2 (panel a) and $n$ + $^{40}$Ca; $E_{n}$=20.43 MeV for LE (19.13 MeV for NL), $\ell$=2 and $j$ =5/2 (panel b) . We show the results obtained including nonlocality in $U_{An}$ (NL, solid line) and those obtained using the corresponding LEP (LE,  dotted line). It can be seen  that nonlocality reduces the diagonal part of the Green's function in the internal region with respect to the LE calculation, and shifts the function at larger distances. 
		
		\begin{figure}
		\includegraphics[width=.45\textwidth,height=6.8cm]{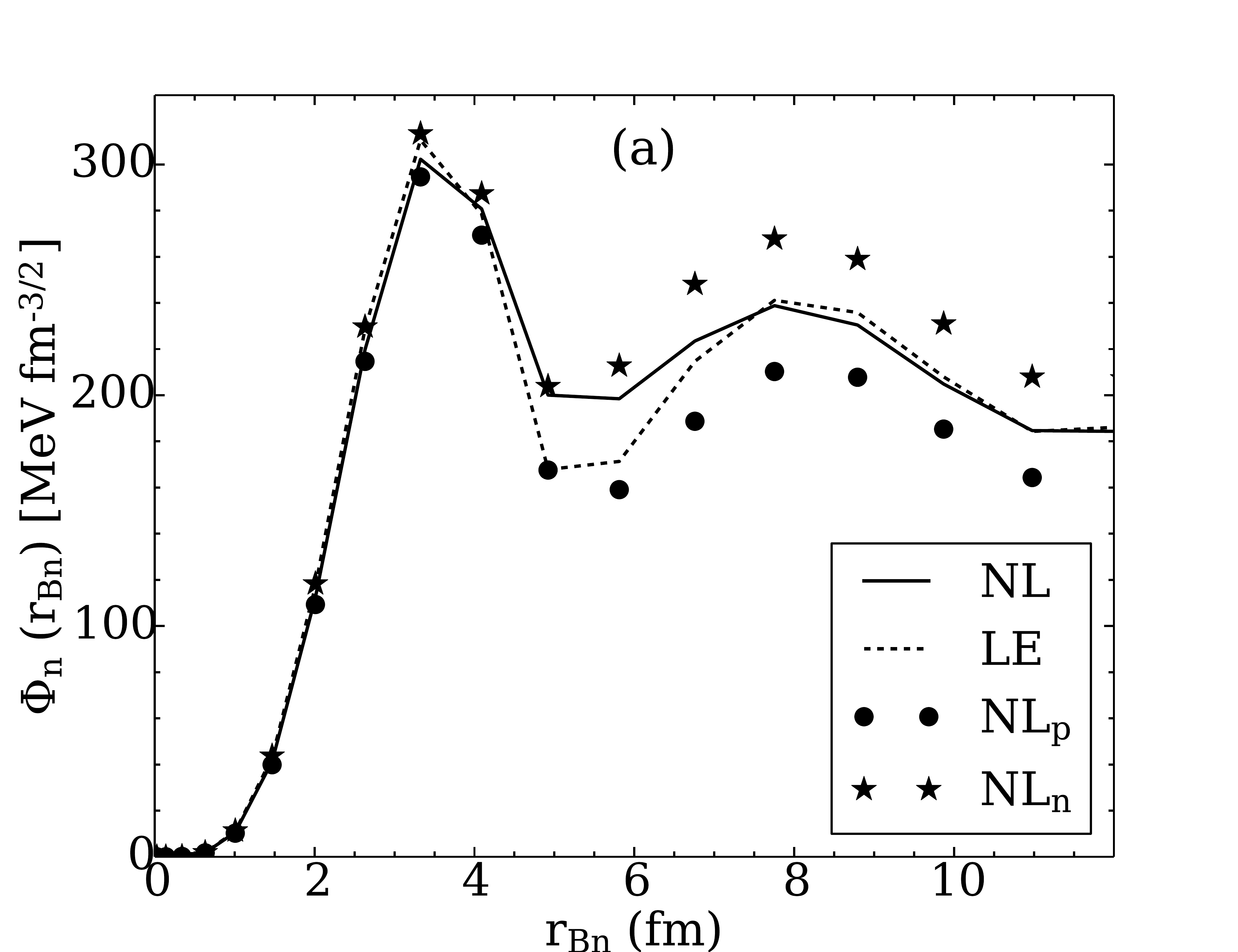}\hfill%
		\includegraphics[width=.45\textwidth,height=6.8cm]{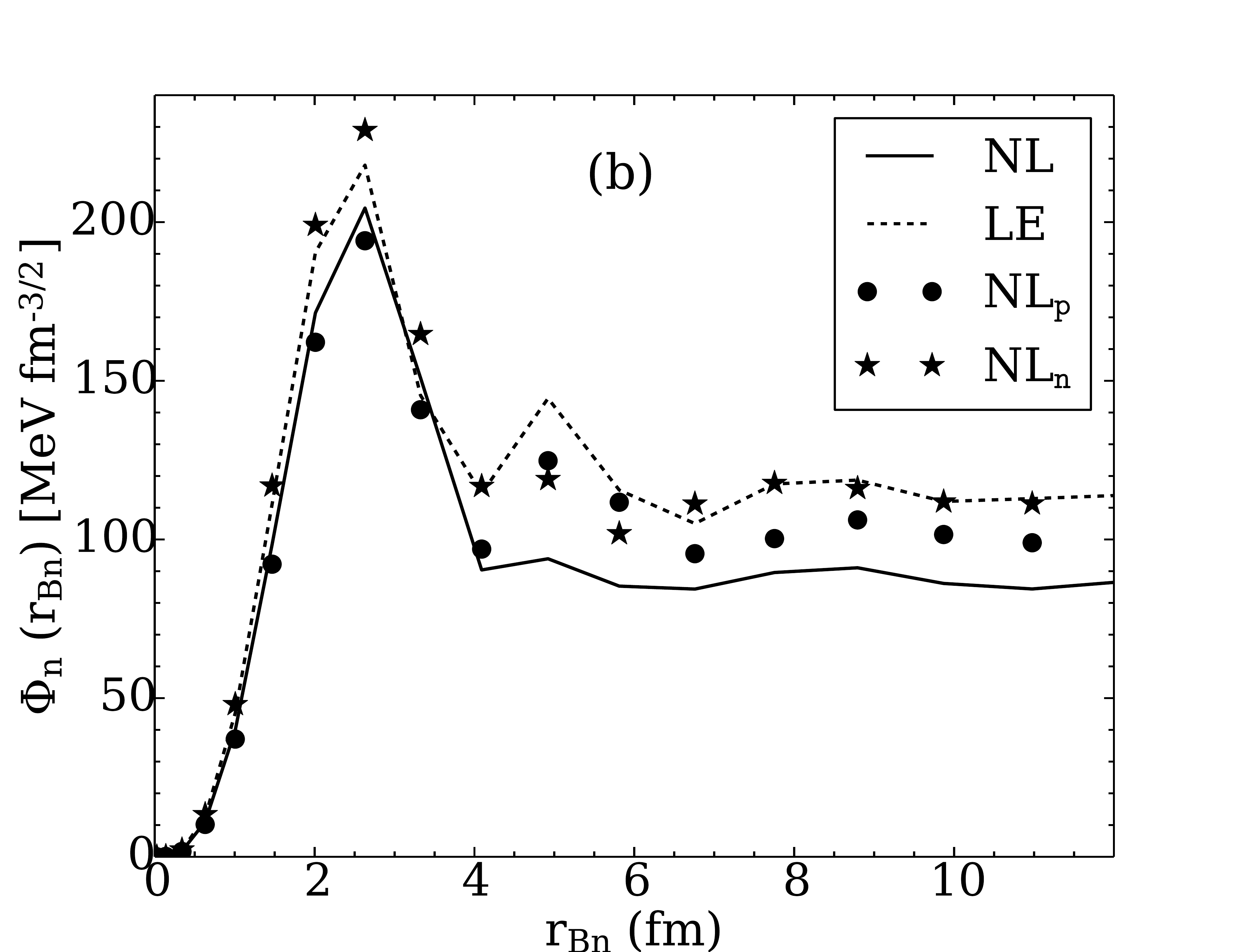}\hfill%
		\caption{Neutron-$^{40}$Ca wave function (see Eq. (\ref{equ:neutronwave})) as a function of radius, for: (a) $\ell$=3 and $j$=7/2, $E_{p}$=10.5 MeV for LE (10.9 MeV for NL$_{\text{n}}$, NL$_{\text{p}}$ and NL) and panel (b) $\ell$=2, and $j$=5/2, $E_{p}$=26.1 MeV for LE ( 27.2 MeV for NL$_{\text{n}}$, 26.8 MeV for NL$_{\text{p}}$, and 27.4 MeV for NL). We show the results for the different calculations discussed in the text.} 
		\label{fig:neutronwave} 
		\end{figure}
		
The neutron wavefunction in the field of the target  following the breakup of the deuteron is a product of the source $S$ and the Green's function $G_B^{\text{opt}}$(Eq.~(\ref{equ:neutronwave})).  The results corresponding to the cases considered in Figs. \ref{fig:source} and \ref{fig:green} are displayed in Fig.~\ref{fig:neutronwave}.
By comparing the results with local interactions (dotted line) with those using non--local interactions (solid line) we find that overall nonlocality reduces the wavefunction in the interior but its effect in the exterior region depends on the beam energy, caused by the radial shift in the Green's function shown in Fig.~\ref{fig:green}.    Fig.~\ref{fig:neutronwave} also shows that nonlocality in $U_{An}$ and $U_{Ap}$ have opposite effects: while the  neutron wavefunction obtained including nonlocality in $U_{An}$ only  (NL$_{\text{n}}$, stars) is larger than the local counterpart (LE), the neutron wavefunction obtained including nonlocality in $U_{Ap}$ only  (NL$_{\text{p}}$, circles) is smaller than the local one.

\begin{figure}[p]
			\includegraphics[width=.45\textwidth,height=6.9cm]{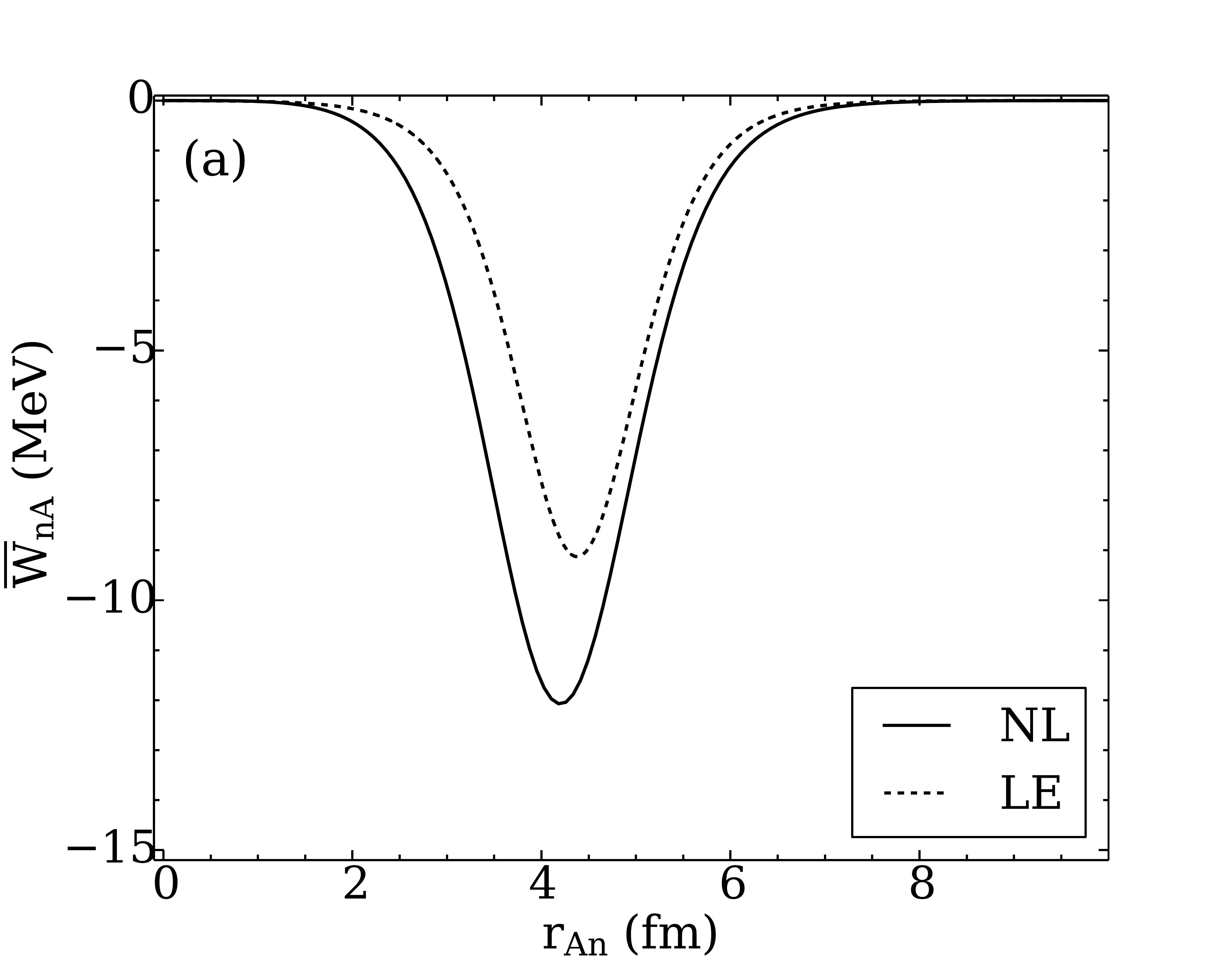}\hfill%
			\includegraphics[width=.45\textwidth,height=6.7cm]{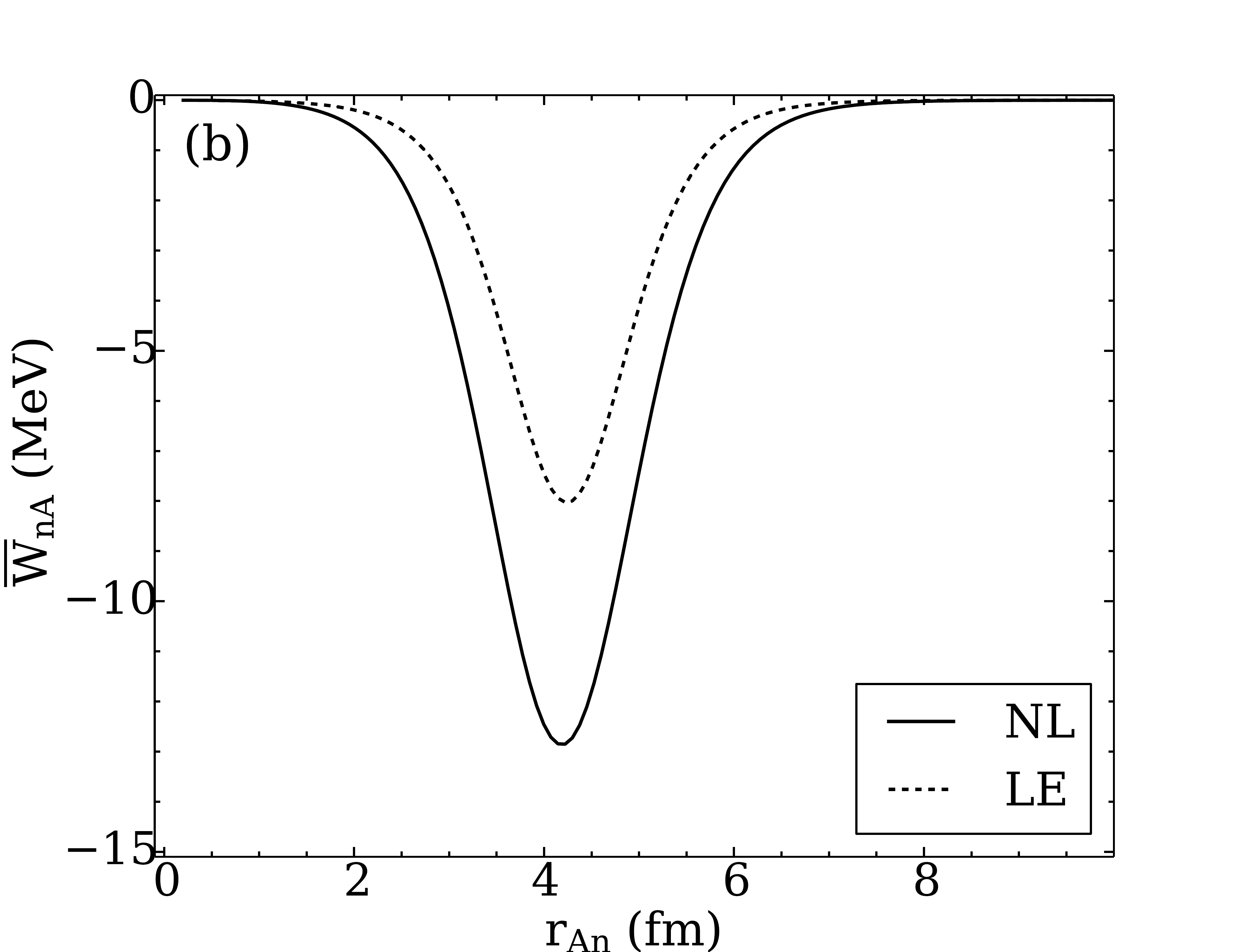}\hfill%
			\caption{We illustrate in this figure the differences between the imaginary parts of the NL and LE potentials. For the NL potential, we plot the quantity $\int W_{An}(r_{An},r_{An}') r_{An}'^2 dr_{An}'$. The two cases shown are: (a) n+$^{40}$Ca at $E_{n}$=6.74 MeV  and (b) n+$^{40}$Ca at $E_{n}$=19.13 MeV.} 
			\label{fig:potential} 
			\end{figure}

Finally, we look at the imaginary term $W_{An}$. In order to be able to compare the non--local potential with the local equivalent (Table \ref{tab:table_parameters}), we plot  the  non--local neutron potential after integration over one of the radial variables $(\bar W_{An}(r)=\int W_{An}(r,r')r'^{2}dr')$.  We pick the same cases relevant to the examples shown in Fig. \ref{fig:source}. In Fig.~\ref{fig:potential} we show the imaginary potential for the $^{40}$Ca+$n$ system for: $E_{n}$=6.34 MeV (panel a) and $E_n$=19.13 MeV (panel b) (these energies correspond to the neutron energies at the peak of the NEB energy distribution for $(d,p)$ at 20 MeV and 50 MeV respectively). It is apparent that the absolute value of the imaginary term of the non--local potential is larger than the one of the corresponding local equivalent. 

By inspecting the various ingredients of the  reaction theory, we are now able to justify the varied effects that nonlocality can have on the NEB cross sections. The overall reduction of the neutron wavefunction $\Phi_n$  in the interior competes with the increase of the magnitude of the non--local interaction $W_{An}$ to produce the complex effects on the cross sections discussed in Section IV.

\section{Conclusion}

In this work, we explore the effects of nonlocality in $A(d,p) B^{*}$ reactions populating continuum states in the final nucleus within the model described in ~\cite{Potel_PRC2015}.
This study is an extension of the study done in ~\cite{Titus_prc2014} for $A(d,p)B$ to bound states. 

This is a systematic study: we include $(d,p)$ reactions on $^{16}$O, $^{40}$Ca, $^{48}$Ca and $^{208}$Pb at varies beam energies $E_{d}$=10 MeV, 20 MeV and 50 MeV. We analyse energy distributions,  angular distributions as well as spin distributions. We compare results obtained using non--local $U_{An}$ and $U_{Ap}$ interactions, with those obtained using the corresponding local equivalent potentials.

Our results demonstrate that the importance of the effect of nonlocality on the magnitude and the shape of the distributions is dependent on beam energy, target, and final spin state. Overall, we find that nonlocality has a non-negligible effect on the non-elastic breakup cross sections, the component that is relevant to the surrogate method and to the extraction of the neutron capture cross section. The effect can be as large as $\sim$ 85\% in some cases, especially for heavier targets.

Understanding the specific origins of the overall effect on the cross sections is not easy, due to the complexity of the reaction theory used to describe these inclusive processes. Nonlocality acts in such a way that it produces a reduction in the interior of the  neutron wavefunction in the field of the target, following the breakup.
At the same time, the imaginary term of the neutron optical  potential, responsible for the {\it capture} of the neutron into the final state $B^{*}$ increases in absolute value.
These two effects compete and can produce large variations on the resulting cross section, including both enhancements and reductions.

Nevertheless, and even if the magnitude of the cross sections can change considerably, the relative weights of the various spin contributions are mostly unaltered with nonlocality. Since this is the most important input to the surrogate method analysis, it is reassuring that previous studies will not be brought into question with the results presented here~\cite{Potel_EPJA2017,Potel_PRC2018,Potel_PRC2015}.
		
\section*{Acknowledgement}

We gratefully acknowledge I. J. Thompson for useful discussions and comments.  This work was supported by the National Science
	Foundation under Grant  PHY-1520929, the U.S. Department of Energy National Nuclear Security Administration under the Stewardship Science Academic Alliances program, NNSA Grant No. DE-NA0000979 and No. DE-FG52-
	08NA28552.  

 \bibliographystyle{apsrev4-1}
 \bibliographystyle{ieeetr}
 \bibliography{nonlocal.bib}

\end{CJK}

\end{document}